\documentclass[a4paper,11pt]{article}
\pdfoutput=1 

\usepackage{jcappub} 

\usepackage[T1]{fontenc} 
\usepackage{placeins}
\usepackage{subcaption}
\usepackage{cleveref}
\usepackage{stfloats}
\usepackage{tabularx}
\usepackage{array}

\title{\boldmath Fast and accurate large-volume Lyman-$\alpha$ forest mocks with three-dimensional U-Net trained on IllustrisTNG and FLAMINGO}

\author[a,1]{Bhaskar Arya,\note{Corresponding author.}}
\author[b]{Soumak Maitra}
\author[b]{and Shadab Alam}

\affiliation[a]{Department of Astronomy and Astrophysics,\\Raman Research Institute,\\ Bengaluru, India}
\affiliation[b]{Department of Theoretical Physics,\\
Tata Institute of Fundamental Research,\\Mumbai, India}

\emailAdd{bhaskararya4@gmail.com}
\emailAdd{soumak.maitra@theory.tifr.res.in}
\emailAdd{shadab@theory.tifr.res.in}

\abstract{The Lyman-$\alpha$ forest probes matter structure at $z\gtrsim2$, constraining baryon acoustic oscillations, neutrino masses, and dark matter properties. Interpreting current and upcoming spectroscopic surveys requires mock catalogues spanning Gpc-scale volumes while capturing small-scale intergalactic medium physics. This combination of high resolution and larger volume is prohibitively expensive for hydrodynamical simulations. We present a field-level emulator, based on a three-dimensional U-Net, that maps approximate density and velocity fields directly to Ly$\alpha$ optical depth and transmitted flux. At $z\sim3$, we train in small volumes and test on much larger, independent ones in two settings: IllustrisTNG (training on a $110\,\mathrm{cMpc}$ box, testing on a $300\,\mathrm{cMpc}$ box with 2LPT and $N$-body inputs) and FLAMINGO (training on $1\,\mathrm{cGpc}$, testing on $2.8\,\mathrm{cGpc}$ with $N$-body inputs).The hydrodynamical statistics of TNG and FLAMINGO differ appreciably over their common \(k\)-range, by \(\sim 15\%\) in \(P_{\rm 1D}\), \(\sim 17\%\) in the bispectrum, and \(\sim 40\%\) in the connected trispectrum, although their normalized power-spectrum correlation structures remain broadly similar. Within each suite, the emulator reproduces \(P_{\rm 1D}\) to \(2.8\%\) in TNG and \(5.6\%\) in FLAMINGO, the bispectrum to \(4\)–\(5\%\), and the trispectrum to \(10\)–\(16\%\), while also recovering the flux PDF and broad correlation structure. Field-level emulation thus provides a practical route to large-volume Ly\(\alpha\)-forest mocks for current and future surveys.}

\begin{document}
\maketitle
\flushbottom

\section{Introduction}

The Lyman-$\alpha$ forest --- the absorption features imprinted on quasar spectra by intervening neutral hydrogen in the intergalactic medium (IGM) --- maps the cosmic web in one dimension along each line of sight, making it one of the most powerful probes of large-scale structure at $z\sim2$--$6$ \citep{Rauch1998, Croft2002, McDonald2006, Meiksin2009, PalanqueDelabrouille2013, McQuinn2016, Chabanier2019, Bosman2022}. On mildly non-linear scales the forest traces the matter density, while on smaller scales it responds to pressure smoothing, thermal broadening, peculiar velocities, and the ultraviolet background. It therefore constrains the matter power spectrum \citep{Croft2002, McDonald2006, PalanqueDelabrouille2013, Chabanier2019}, neutrino masses \citep{Seljak2006, Viel2010, PalanqueDelabrouille2015, Yeche2017},
dark-matter models \citep{Viel2013, Baur2016, Irsic2017WDM, Irsic2017FDM,
Murgia2018, RogersPeiris2021, Baur2017Sterile}, the thermal history of the IGM \citep{Schaye2000, Becker2011, Boera2014, Walther2019, Villasenor2021}, and reionization \citep{Worseck2011, Worseck2016, Boera2019}. Surveys such as SDSS-BOSS \citep{York2000, Dawson2013, Dawson2016}, DESI \citep{DESI2016, DESI2019, DESI2023EDR}, and WEAVE \citep{Pieri2016, Jin2024} further exploit forest correlations to measure baryon acoustic oscillations (BAO) and redshift-space distortions at epochs inaccessible to galaxy surveys.

Interpreting these data requires mock skewers that connect theory to data and quantify survey-level effects such as noise, resolution, continuum fitting, and covariance. Such mocks must cover Gpc-scale volumes while retaining small-scale accuracy along individual lines of sight --- a combination that is computationally prohibitive for hydrodynamical simulations \citep[e.g.][]{Pillepich2018, Nelson2019, Bolton2017, Lukic2015, Schaye2015, Schaye2023, Chabanier2024}, even before the many realizations needed for covariance estimation are considered.

Approximate methods reduce this cost. Lognormal models \citep{BiDavidsen1997, Choudhury2001a, Choudhury2001b, Arya2023, Arya2024a, Arya2024b} and fluctuating Gunn--Peterson schemes \citep{Kooistra2022, Sinigaglia2024, Etourneau2024} underpin end-to-end survey-mock pipelines \citep{Bautista2015, Farr2020}, but rely on simplified prescriptions for the density field, thermal state, and density--flux relation. Calibration-based methods instead map dark matter to flux using a reference hydrodynamical simulation --- LyMAS and LyMAS2 sample conditional flux distributions given smoothed density and velocity fields \citep{Peirani2014, Peirani2022}, while Hydro-BAM learns a three-dimensional bias relation that also recovers higher-order statistics \citep{Sinigaglia2022} --- but are designed to reproduce selected summary statistics rather than the flux along individual skewers.

Deep learning enables a more flexible, field-level mapping. \cite{Harrington:2021srm} trained a convolutional network to reconstruct baryon density, temperature, and velocity from $N$-body fields in Nyx \citep{Lukic2015}, HyPhy introduced a probabilistic variant based on a conditional variational autoencoder \citep{Horowitz2021olb}, and LyAl-Net used a U-Net to emulate neutral hydrogen density and temperature, porting the model across simulations with different gas physics \citep{Boonkongkird2023}. A complementary approach enhances low-resolution hydrodynamical simulations directly, and has been used to reconstruct a Gpc-scale volume from an $80\,h^{-1}\,\mathrm{Mpc}$ training box \citep{Jacobus2023, Jacobus2025}. These studies demonstrate that neural networks can recover the Ly$\alpha$ forest at the field level. Because the networks are fully convolutional, they can in principle be trained on small boxes and deployed on much larger volumes \citep{Harrington:2021srm} --- but this transfer has not been validated directly against a hydrodynamical simulation of the target volume at Gpc scales, nor beyond two-point statistics, for gravity-only inputs.

In this work we close this gap. We train an anisotropic three-dimensional U-Net to predict the Ly$\alpha$ optical depth, and hence transmitted flux, directly from approximate dark matter density and velocity fields, so that the model learns the non-linear, non-local mapping in a small calibration volume and can then be applied to volumes far beyond the reach of hydrodynamical simulations. We test the transfer in two regimes at $z\sim3$: with IllustrisTNG, training on a $110\,\mathrm{cMpc}$ box and testing on an independent $300\,\mathrm{cMpc}$ box, using both second-order Lagrangian perturbation theory (2LPT) and $N$-body inputs to isolate the impact of the gravitational field quality; and with FLAMINGO, training on a $1\,\mathrm{cGpc}$ box and testing on a $2.8\,\mathrm{cGpc}$ volume with $N$-body inputs. In both cases the test volume is roughly twenty times larger than the training volume. Against the corresponding hydrodynamical simulations, we validate the emulated flux using the probability distribution function, the one-dimensional flux power spectrum, bispectrum and trispectrum, and the power spectrum covariance --- statistics that existing calibration-based mocks target but field-level emulators have not yet been shown to reproduce after cross-volume transfer.

The paper is organized as follows. We describe the U-Net architecture in \cref{sec:network_architecture} and the training and testing setup in \cref{sec:methodology}. We present our results in \cref{sec:results}, discuss implications and limitations in \cref{sec:discussion}, and summarize in \cref{sec:conclusion}.

\section{Methodology}
\label{sec:methodology}

We train three-dimensional U-Nets to map gravity-only density and line-of-sight velocity fields to the Ly$\alpha$ optical-depth field of a hydrodynamical simulation. Our principal aim is to determine whether a mapping calibrated in one volume remains accurate when applied to a substantially larger, independent volume. We test this separately in the IllustrisTNG and FLAMINGO simulation suites, using fields at $z\sim3$. The simulation volumes, input fields, and training and testing samples are summarized in \cref{tab:setup}.

\subsection{Simulations}
\label{subsec:simulations}

The two simulation suites provide complementary tests. IllustrisTNG \citep{Pillepich2018,Nelson2019} is evolved with the moving-mesh code \textsc{arepo} \citep{Springel2010}, whereas FLAMINGO \citep{Schaye2023} uses the smoothed-particle-hydrodynamics code \textsc{swift} \citep{Schaller2024}. The suites also adopt different cosmologies, feedback prescriptions, and ultraviolet backgrounds, and FLAMINGO evolves massive neutrinos as a separate species. Consequently, the relation between density, velocity, and Ly$\alpha$ optical depth differs between the two suites.

For IllustrisTNG, we use a $110\,\mathrm{cMpc}$ box with $455^3$ particles for training and a $300\,\mathrm{cMpc}$ box with $1250^3$ particles for testing. The fields are represented on $448^3$ and $1248^3$ meshes, respectively, with a common cell size of approximately $0.24\,\mathrm{cMpc}$. For FLAMINGO, we use a $1\,\mathrm{cGpc}$ box with $1800^3$ particles for training and a $2.8\,\mathrm{cGpc}$ box with $5040^3$ particles for testing. These fields are represented on $1800^3$ and $5040^3$ meshes with a common cell size of approximately $0.56\,\mathrm{cMpc}$.

FLAMINGO therefore reaches the Gpc-scale volumes relevant for surveys, although at a coarser spatial resolution than IllustrisTNG. Within each suite, the training and test boxes have the same cell size, while the test volume is approximately twenty times larger than the training volume. Each experiment consequently isolates transfer in simulation volume at fixed spatial resolution.

\begin{table}
    \centering
    \caption{Summary of the simulation volumes and samples used for training, validation, and testing. All fields are evaluated at $z\sim3$.}
    \label{tab:setup}
    \begin{tabular}{lcc}
        \hline
         & IllustrisTNG & FLAMINGO \\
        \hline
        Hydrodynamics code & \textsc{arepo} & \textsc{swift} \\
        Training box & $110\,\mathrm{cMpc}$, $448^3$ & $1\,\mathrm{cGpc}$, $1800^3$ \\
        Test box & $300\,\mathrm{cMpc}$, $1248^3$ & $2.8\,\mathrm{cGpc}$, $5040^3$ \\
        Cell size & $0.24\,\mathrm{cMpc}$ & $0.56\,\mathrm{cMpc}$ \\
        Test/training volume & $\simeq20$ & $\simeq22$ \\
        Inputs & 2LPT, $N$-body & $N$-body \\
        Training samples & $4000\times(8\times8\times448)$ & $400\times(8\times8\times1800)$ \\
        Validation & $112\times112\times448$ & $100\times(8\times8\times1800)$ \\
        Test samples & $5000\times(8\times8\times1248)$ & $1500\times(8\times8\times5040)$ \\
        \hline
    \end{tabular}
\end{table}

\subsection{Input and target fields}
\label{subsec:fields}

The target is the hydrodynamical Ly$\alpha$ optical-depth field, $\tau$, from which the transmitted flux is calculated as
\begin{equation}
    F=\exp(-\tau).
\end{equation}
The network inputs are the density and line-of-sight velocity fields obtained from a gravity-only description of the same volume. The input and target fields are sampled on the same grid so that their skewers are spatially aligned.

We consider two descriptions of the gravity-only fields. Second-order Lagrangian perturbation theory (2LPT) provides an inexpensive approximation that captures large-scale modes but not the fully non-linear evolution. $N$-body simulations provide a more accurate description of the non-linear density and velocity fields while remaining considerably less expensive than hydrodynamical simulations. For IllustrisTNG, we train separate networks using 2LPT and $N$-body inputs against the same hydrodynamical target. For FLAMINGO, we use $N$-body inputs only.

\subsection{Neural-network model}
\label{sec:network_architecture}

\subsubsection{Architecture}

We use an anisotropic three-dimensional U-Net to predict $\log_{10}\tau$ from the gravity-only density and line-of-sight velocity fields. U-Net was originally introduced by \cite{Ronneberger2015} for biomedical image segmentation and was subsequently extended to volumetric data using three-dimensional convolutions \citep{Cicek2016}. Its encoder--decoder structure combines an extended receptive field with skip connections that preserve localized information, making it well suited to field-level inference on cosmological grids. U-Net-like architectures have also been applied to Ly$\alpha$-forest tomography; for example, \cite{Maitra2025} used a three-dimensional variational autoencoder with a U-Net architecture to reconstruct the dark-matter density field from Ly$\alpha$-forest spectra.

The network receives two input channels containing the density and line-of-sight velocity fields, with dimensions $2\times D\times8\times8$. The dimension $D$ follows the skewer direction, while the two transverse dimensions retain information from neighbouring sightlines. The predicted optical depth along a skewer can therefore depend on both its line-of-sight structure and its surrounding three-dimensional environment.

Downsampling is applied only along the line-of-sight direction, using max pooling with kernel and stride $(2,1,1)$, so the $8\times8$ transverse dimensions remain unchanged. The TNG network contains four encoder levels with $8$, $16$, $32$, and $64$ feature channels, followed by a $192$-channel bottleneck. The FLAMINGO network contains three encoder levels with $8$, $16$, and $32$ channels, followed by a $96$-channel bottleneck. The decoder reverses these channel sequences and restores the line-of-sight resolution using transposed convolutions with kernel and stride $(2,1,1)$. At each level, an additive skip connection combines the upsampled decoder features with the corresponding encoder features.

Each encoder, bottleneck, and decoder block contains two $3\times3\times3$ convolutions followed by ReLU activations, with dropout of $p=0.02$ between the convolutions. Circular padding is used in all directions. The encoder convolutions employ line-of-sight dilation factors $(1,2,2,4)$ for TNG and $(1,2,2)$ for FLAMINGO, with corresponding bottleneck dilation factors of $8$ and $4$. A final $1\times1\times1$ convolution maps the decoded eight-channel representation to a single $\log_{10}\tau$ field. The transmitted flux is then obtained as
\begin{equation}
    F=\exp\!\left(-10^{\log_{10}\tau}\right).
\end{equation}

\begin{figure}
    \centering
    \includegraphics[width=\linewidth,height=0.5\linewidth]{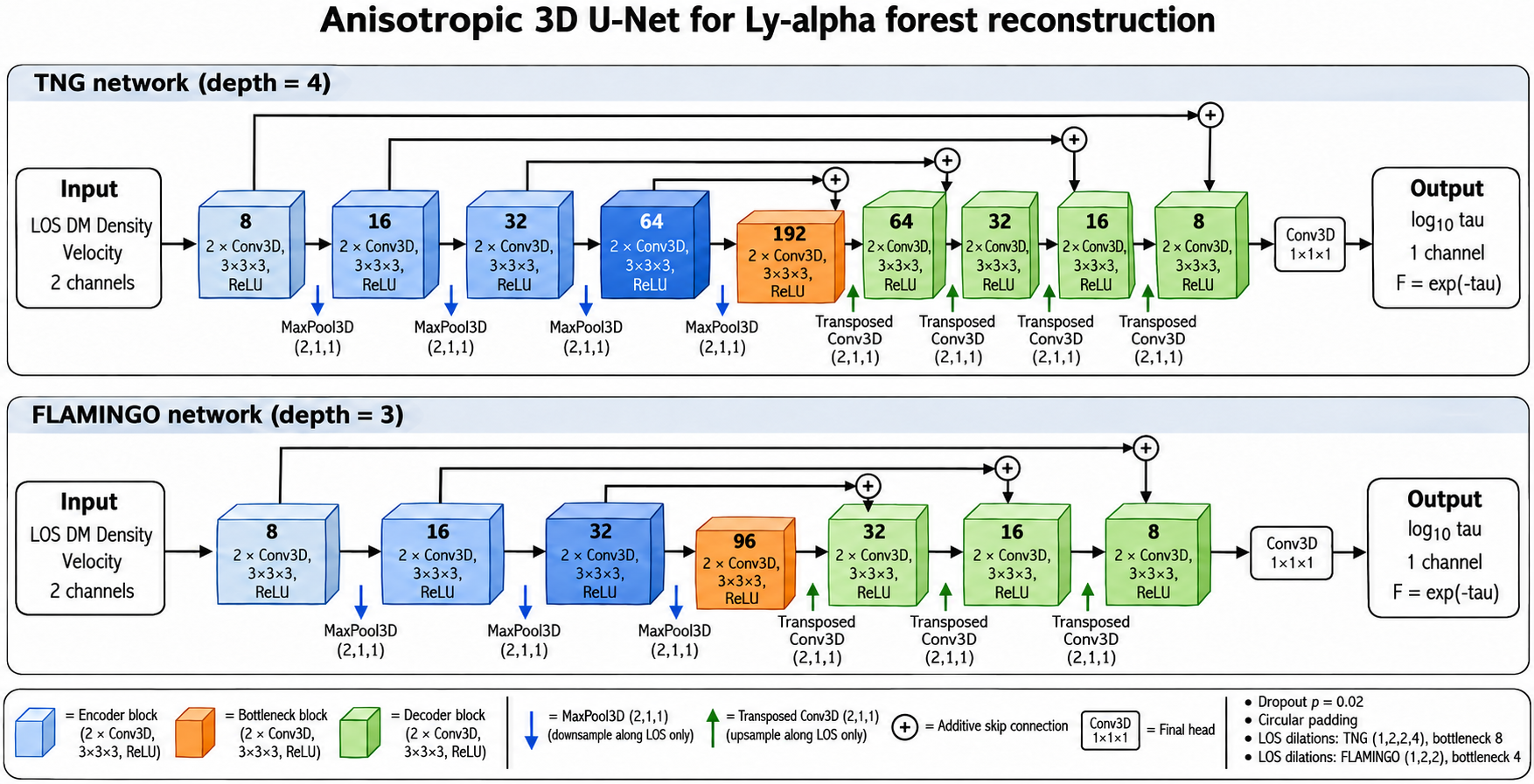}
    \caption{Schematic representation of the anisotropic three-dimensional U-Net architectures used in this work. The two input channels contain the line-of-sight dark-matter density and velocity fields. Each convolutional block contains two $3\times3\times3$ convolutions with ReLU activations, while downsampling and upsampling are performed only along the line-of-sight direction using kernels $(2,1,1)$. The TNG and FLAMINGO models use four and three encoder levels, respectively, with bottleneck widths of 192 and 96 channels. Additive skip connections combine the encoder features with the corresponding upsampled decoder features. A final $1\times1\times1$ convolution produces a single $\log_{10}\tau$ field, from which the transmitted flux is obtained as $F=\exp(-\tau)$.}
    \label{fig:unet_architecture}
\end{figure}

\subsubsection{Training}
\label{subsec:training}

Training samples are sub-volumes of \(8\times8\times N_{\rm los}\) cells drawn randomly from the training simulation, where \(N_{\rm los}\) is the complete line-of-sight extent: \(448\) cells for IllustrisTNG and \(1800\) cells for FLAMINGO. We use \(4000\) training samples for IllustrisTNG and \(400\) for FLAMINGO. Validation data are drawn from spatially disjoint regions and comprise a \(112\times112\times448\) block for IllustrisTNG and \(100\) sub-volumes of \(8\times8\times1800\) cells for FLAMINGO.

The networks are optimized with Adam using an initial learning rate of \(5\times10^{-4}\). We use batch sizes of four for TNG and one for FLAMINGO. The learning rate is reduced by a factor of two when the validation loss does not improve for ten epochs, and training is stopped once the validation loss no longer shows sustained improvement. Training requires approximately \(2\,\mathrm{GPU\,h}\) for each simulation suite.

We do not base the reported results on a single checkpoint corresponding to the minimum validation loss. Instead, we retain ten checkpoints from the converged region around this minimum and evaluate the test volume with each of them. The central prediction is the median over these checkpoints, while the 16th--84th percentile range quantifies the systematic sensitivity to stochastic optimization and checkpoint selection across training iterations. This spread is not interpreted as a statistical uncertainty associated with the finite simulation volume.

The primary field-level loss is the voxel-wise mean-squared error (MSE) in $\log_{10}\tau$,
\begin{equation}
    \mathcal{L}_{\mathrm{MSE}}
    =
    \frac{1}{N_{\mathrm{vox}}}
    \sum_{i=1}^{N_{\mathrm{vox}}}
    \left[
    \log_{10}\tau_{\mathrm{pred},i}
    -
    \log_{10}\tau_{\mathrm{true},i}
    \right]^2.
    \label{eq:loss_function}
\end{equation}
We supplement this term with a loss that penalizes differences between the predicted and true one-dimensional flux power spectra. For each training batch, the predicted and true optical-depth fields are converted to transmitted flux, and $P_{\rm 1D}(k)$ is computed along the line-of-sight direction. The power-spectrum loss is evaluated over every non-zero native Fourier mode, from the fundamental mode $k_{\rm fund}=2\pi/L$ to the Nyquist frequency $k_{\rm Nyq}$, without additional $k$-binning:
\begin{equation}
    \mathcal{L}_{\mathrm{PS}}
    =
    \frac{1}{N_k}
    \sum_{n=1}^{N_k}
    \left[
    \log P_{\rm 1D}^{\rm pred}(k_n)
    -
    \log P_{\rm 1D}^{\rm true}(k_n)
    \right]^2,
    \label{eq:ps_loss}
\end{equation}
where $N_k$ is the number of included Fourier modes.

Two auxiliary prediction heads at intermediate decoder levels provide additional MSE supervision. Denoting their losses by $\mathcal{L}_{\mathrm{aux},1}$ and $\mathcal{L}_{\mathrm{aux},2}$, the complete training objective is
\begin{equation}
    \mathcal{L}_{\mathrm{tot}}
    =
    \mathcal{L}_{\mathrm{MSE}}
    +
    \lambda_{\mathrm{PS}}\mathcal{L}_{\mathrm{PS}}
    +
    \lambda_{\mathrm{aux},1}\mathcal{L}_{\mathrm{aux},1}
    +
    \lambda_{\mathrm{aux},2}\mathcal{L}_{\mathrm{aux},2},
    \label{eq:total_loss}
\end{equation}
with $\lambda_{\mathrm{PS}}=0.05$, $\lambda_{\mathrm{aux},1}=0.05$, and $\lambda_{\mathrm{aux},2}=0.02$. The MSE term drives the field-level reconstruction but tends to smooth small-scale fluctuations and suppress high-$k$ power when used alone. The power-spectrum term counteracts this tendency by directly constraining scale-dependent clustering, while the auxiliary heads improve optimization at intermediate decoder depths.

\subsubsection{Application to the test volumes}
\label{subsec:testing}

After training, each network is applied to the independent test box from the same simulation suite, with its weights held fixed. For IllustrisTNG, we evaluate $5000$ randomly selected sub-volumes of $8\times8\times1248$ cells. For FLAMINGO, we evaluate $1500$ sub-volumes of $8\times8\times5040$ cells. The hydrodynamical flux in each test volume is used only as the reference against which the predictions are evaluated.

Because the line-of-sight dimension spans the full length of each test box, these samples contain modes with wavelengths longer than those present in the corresponding training volume. Their recovery therefore directly tests whether the learned density--velocity--optical-depth mapping generalizes beyond the volume used for calibration.

\subsection{Validation statistics and baseline model}
\label{subsec:validation_statistics}

We compare the predicted and hydrodynamical fields at both the field and statistical levels. The field-level comparison uses individual noiseless skewers. The statistical comparison uses the flux probability distribution function, the one-dimensional flux power spectrum $P_{\rm 1D}$, its correlation matrix, and the flux bispectrum and connected trispectrum. These statistics probe, respectively, the distribution of absorption strengths, line-of-sight clustering, scale-to-scale mode coupling, and non-Gaussian correlations beyond the two-point level.

For all Fourier-space statistics, we use the flux contrast
\begin{equation}
    \delta_F(x)=\frac{F(x)}{\langle F\rangle}-1,
    \label{eq:flux_contrast}
\end{equation}
where $\langle F\rangle$ denotes the mean over the full set of skewers used for the corresponding statistic, and $\widetilde{\delta}_F(k)$ denotes its Fourier transform.

The flux probability distribution function \(p(F)\) describes the one-point distribution of transmitted-flux values. For a set of \(N_{\rm pix}\) pixels, it can be written as
\begin{equation}
    p(F)
    =
    \frac{1}{N_{\rm pix}}
    \sum_{i=1}^{N_{\rm pix}}
    \delta_{\rm D}(F-F_i),
    \label{eq:flux_pdf_definition}
\end{equation}
and is normalized such that
\begin{equation}
    \int_0^1 p(F)\,\mathrm{d}F=1.
\end{equation}
In practice, we estimate \(p(F)\) by binning all pixel fluxes from the relevant set of skewers. The flux PDF therefore tests whether the predicted field reproduces the relative abundance of pixels with different absorption strengths.

The one-dimensional flux power spectrum quantifies two-point clustering along the line-of-sight direction. With the Fourier convention used here, it is defined by
\begin{equation}
    \left\langle
    \widetilde{\delta}_F(k)
    \widetilde{\delta}_F(k')
    \right\rangle
    =
    2\pi\,\delta_{\rm D}(k+k')\,
    P_{\rm 1D}(k),
    \label{eq:flux_p1d_definition}
\end{equation}
where the average is taken over the ensemble of skewers. We estimate \(P_{\rm 1D}\) by Fourier transforming \(\delta_F\) along each skewer, computing its squared Fourier amplitude, and averaging over all skewers.

To quantify the coupling between different power-spectrum bins, we first estimate the covariance matrix
\begin{equation}
    C_{ij}
    =
    \frac{1}{N_{\rm s}-1}
    \sum_{a=1}^{N_{\rm s}}
    \left[
    P_{\rm 1D}^{(a)}(k_i)-\overline{P}_{\rm 1D}(k_i)
    \right]
    \left[
    P_{\rm 1D}^{(a)}(k_j)-\overline{P}_{\rm 1D}(k_j)
    \right],
    \label{eq:p1d_covariance_definition}
\end{equation}
where \(P_{\rm 1D}^{(a)}(k_i)\) is the power measured from sample \(a\) in bin \(k_i\), \(N_{\rm s}\) is the number of samples, and
\begin{equation}
    \overline{P}_{\rm 1D}(k_i)
    =
    \frac{1}{N_{\rm s}}
    \sum_{a=1}^{N_{\rm s}}
    P_{\rm 1D}^{(a)}(k_i)
\end{equation}
is the sample-averaged power spectrum. The corresponding correlation matrix is
\begin{equation}
    r_{ij}
    =
    \frac{C_{ij}}
    {\sqrt{C_{ii}C_{jj}}},
    \label{eq:p1d_correlation_definition}
\end{equation}
where \(r_{ij}\in[-1,1]\) and \(r_{ii}=1\). The off-diagonal coefficients measure the strength and sign of the coupling between different \(k\)-bins. Because the covariance is normalized by the variance in each bin, the correlation matrix tests the pattern of mode coupling but not the absolute covariance amplitude.

To measure non-Gaussianities, we define the 1D flux bispectrum through
\begin{equation}
    \left\langle
    \widetilde{\delta}_F(k_1)
    \widetilde{\delta}_F(k_2)
    \widetilde{\delta}_F(k_3)
    \right\rangle
    =
    2\pi\,\delta_{\rm D}(k_1+k_2+k_3)
    B(k_1,k_2,k_3),
    \label{eq:flux_bispectrum_definition}
\end{equation}
where the Dirac delta imposes $k_1+k_2+k_3=0$. Only two wavenumbers are independent in one dimension, and we therefore write
\begin{equation}
    B(k_1,k_2)\equiv B(k_1,k_2,-k_1-k_2).
    \label{eq:flux_bispectrum_1d}
\end{equation}

Similarly, the connected 1D flux trispectrum is defined through
\begin{equation}
    \left\langle
    \widetilde{\delta}_F(k_1)
    \widetilde{\delta}_F(k_2)
    \widetilde{\delta}_F(k_3)
    \widetilde{\delta}_F(k_4)
    \right\rangle_{\rm c}
    =
    2\pi\,\delta_{\rm D}(k_1+k_2+k_3+k_4)
    T(k_1,k_2,k_3,k_4),
    \label{eq:flux_trispectrum_definition}
\end{equation}
where the subscript ${\rm c}$ denotes the connected component. Momentum conservation fixes the fourth mode, so that
\begin{equation}
    T(k_1,k_2,k_3)
    \equiv
    T(k_1,k_2,k_3,-k_1-k_2-k_3).
    \label{eq:flux_trispectrum_1d}
\end{equation}
The connected trispectrum determines the non-Gaussian contribution to the covariance of $P_{\rm 1D}$.

To quantify differences in the positive 1D power spectrum, we use the mean absolute logarithmic ratio
\begin{equation}
    D_{\rm log}
    =
    \left\langle
    \left|
    \ln\!\left(\frac{P_1(k)}{P_2(k)}\right)
    \right|
    \right\rangle_k,
    \label{eq:mean_log_difference}
\end{equation}
which approaches the mean absolute fractional difference for small deviations. Because the bispectrum and connected trispectrum are signed, we compare them using the symmetric normalized RMS difference
\begin{equation}
    D_{\rm NRMS}(S_1,S_2)
    =
    \frac{
    \sqrt{\langle(S_1-S_2)^2\rangle}
    }{
    \sqrt{\tfrac{1}{2}\left[
    \langle S_1^2\rangle+
    \langle S_2^2\rangle
    \right]}
    },
    \label{eq:normalized_rms_difference}
\end{equation}
evaluated over all elements of the common two-dimensional $k$-bin block. For the $P_{\rm 1D}$ correlation matrices, we report the mean absolute off-diagonal difference
\begin{equation}
    D_r
    =
    \left\langle
    \left|r^{(1)}_{ij}-r^{(2)}_{ij}\right|
    \right\rangle_{i<j},
    \label{eq:correlation_absolute_difference}
\end{equation}
excluding the unit diagonal.

\subsubsection{FGPA baseline}

For comparison, we also generate flux fields using the fluctuating Gunn--Peterson approximation (FGPA), which provides a simple and widely used baseline for Ly$\alpha$-forest spectra and mock catalogues \citep{Weinberg1997,FontRibera2012,Etourneau2024,Sinigaglia2024}. For IllustrisTNG, FGPA is applied to the 2LPT density field; for FLAMINGO, it is applied to the $N$-body density field. Assuming a temperature--density slope $\gamma=1.5$, we use
\begin{equation}
    \tau_{\rm FGPA}=A(1+\delta)^\beta,
    \qquad
    \beta=2-0.7(\gamma-1)=1.65,
    \label{eq:fgpa_tau}
\end{equation}
where $\delta$ is the density contrast. The normalization $A$ is calibrated separately in each test volume to reproduce the corresponding hydrodynamical mean transmitted flux.

A concise quantitative summary of the validation results is given in \cref{subsec:results_summary}.

\section{Results}
\label{sec:results}

We compare the network predictions with the hydrodynamical flux in the IllustrisTNG and FLAMINGO test volumes, first skewer by skewer and then through the flux PDF, the 1D power spectrum $P_{\rm 1D}$ and its correlation matrix, and the 1D bispectrum and connected trispectrum. We note that the $P_{\rm 1D}$ enters the training loss through $\mathcal{L}_{\rm PS}$, so its agreement on the test box measures generalisation but is not an independent check. The PDF, the correlation matrices and the higher-order statistics were never targeted during training.

\subsection{Field-level comparison}
\label{subsec:field_level}

\begin{figure}[!htbp]
\centering
\includegraphics[width=\textwidth]{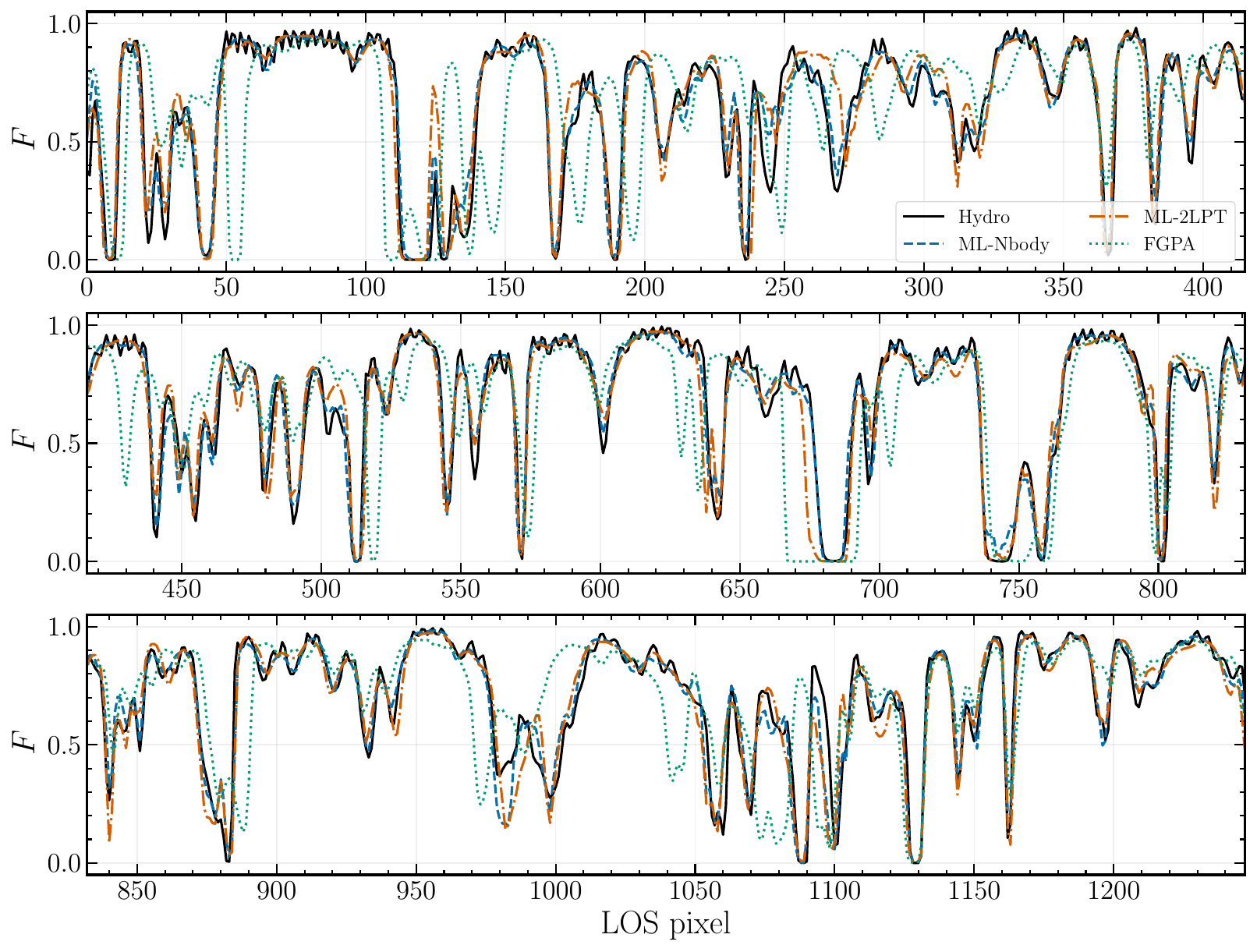}
\caption{Transmitted flux along a randomly chosen skewer in the TNG $300\,\mathrm{cMpc}$ test volume: hydrodynamical reference, network predictions from 2LPT and $N$-body inputs, and FGPA applied to the 2LPT density.}
\label{fig:tng_spectra}
\end{figure}

\Cref{fig:tng_spectra} shows a single TNG skewer. Both networks correctly locate the absorption features and recover their depths and widths for most pixels. By contrast, FGPA fails to reproduce the positions and profiles of individual absorption features because it does not include the effects of peculiar velocities, despite matching the mean transmitted flux by construction. We quantify the field-level agreement using the root-mean-square flux error, \({\rm RMSE}_F=\sqrt{\langle(F_{\rm pred}-F_{\rm hydro})^2\rangle}\), and the Pearson correlation coefficient \(r_F\) between the predicted and hydrodynamical pixel fluxes, where the average is taken over all pixels in the test patches. For the 5000 TNG test patches, the \(N\)-body and 2LPT networks yield \(({\rm RMSE}_F,r_F)=(0.070,0.968)\) and \((0.099,0.937)\), respectively, compared with \((0.287,0.451)\) for FGPA. For the 1500 FLAMINGO test patches, the network yields \((0.147,0.849)\), compared with \((0.263,0.494)\) for FGPA. These metrics confirm that the networks reproduce the hydrodynamical flux at the field level substantially more accurately than FGPA in both simulation suites.

\subsection{Flux PDF and power spectrum}
\label{subsec:pdf_p1d}

\begin{figure*}[!t]
\centering
\includegraphics[width=\textwidth]{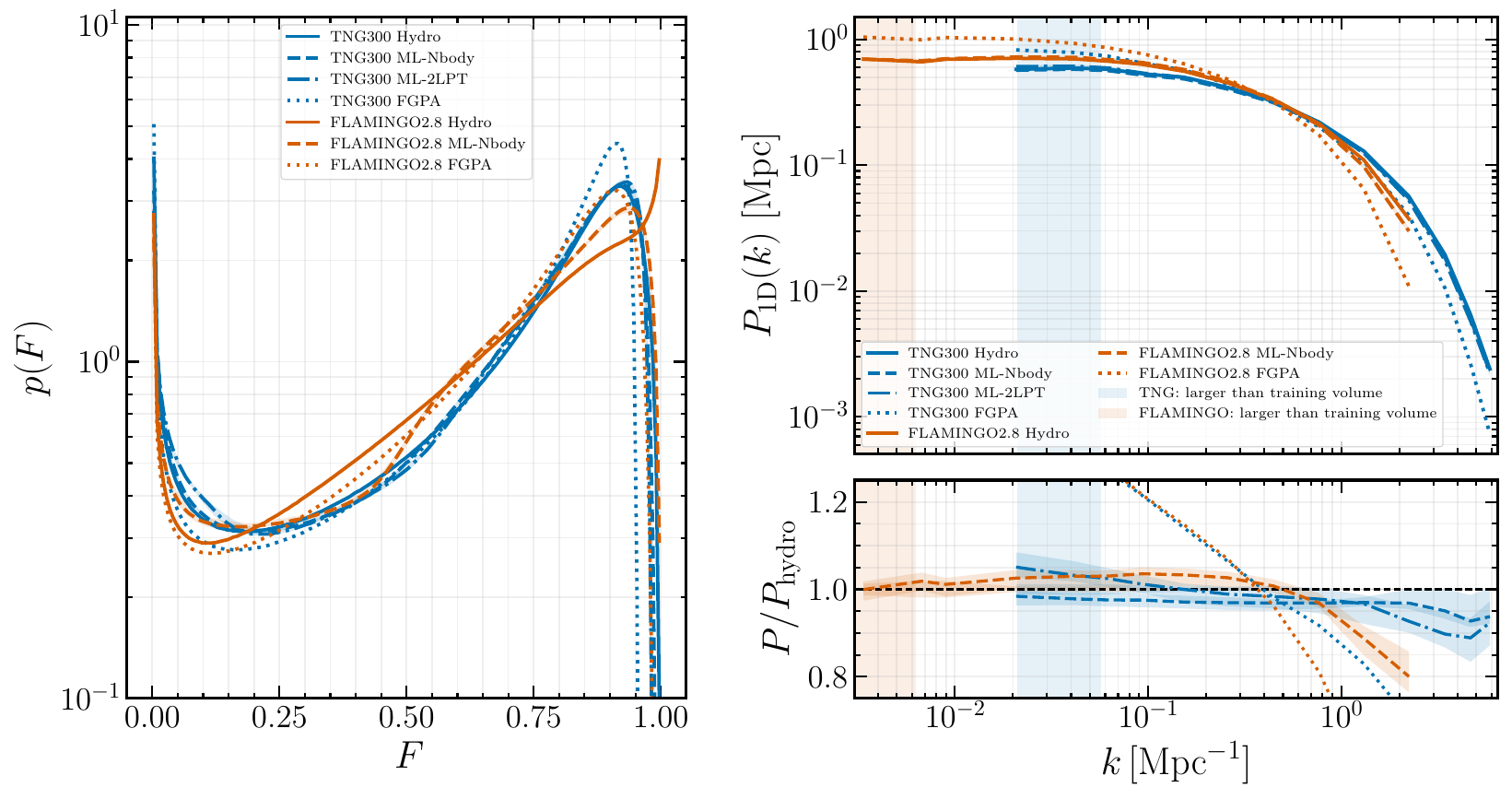}
\caption{Flux PDF (left) and noiseless 1D power spectrum with ratios to the hydrodynamical reference (right) for the TNG and FLAMINGO test volumes. ML curves are medians over ten uniformly sampled post-burn-in checkpoints, with the 16th--84th percentile checkpoint scatter shaded; hydrodynamical bands show the standard error across independent patches. Both statistics are evaluated after mean-flux matching. The blue and orange shaded regions mark $2\pi/300<k<2\pi/110\,\mathrm{Mpc}^{-1}$ for TNG and $2\pi/2800<k<2\pi/1000\,\mathrm{Mpc}^{-1}$ for FLAMINGO, containing two and one bins, respectively. These modes are otherwise not available in training.}
\label{fig:pdf_p1d}
\end{figure*}

\Cref{fig:pdf_p1d} shows the flux PDF and noiseless $P_{\rm 1D}$ for both test volumes. Each predicted field is rescaled to the hydrodynamical mean flux before the power spectrum is computed, so that the comparison isolates scale dependence. Before rescaling, the selected TNG $N$-body and 2LPT models have mean fluxes of 0.691 and 0.687, compared with the hydrodynamical value 0.686, corresponding to offsets of $+0.77\%$ and $+0.15\%$. The selected FLAMINGO model has mean flux 0.696, compared with 0.690 for the hydrodynamical field, an offset of $+0.92\%$. Both the PDFs and $P_{\rm 1D}$ curves in \cref{fig:pdf_p1d} are evaluated after this rescaling. The ML curves are medians over ten checkpoints sampled uniformly from the available post-burn-in epochs, including both endpoints; they are not selected by validation loss. Their 16th--84th percentile bands quantify checkpoint sensitivity and are not statistical errors. The hydrodynamical curves have separate sample-variance bands, computed as $\sigma_P(k)/\sqrt{N_{\rm patch}}$ from 5000 TNG and 1500 FLAMINGO patches. The fractional standard error ranges from $0.14\%$ to $1.20\%$ in TNG and from $0.08\%$ to $1.91\%$ in FLAMINGO.

The headline result is that accuracy does not degrade with volume. For $k\le1\,\mathrm{Mpc}^{-1}$, the networks match the hydrodynamical $P_{\rm 1D}$ to 2.7\% (TNG, $N$-body), 2.1\% (TNG, 2LPT) and 2.2\% (FLAMINGO), measured by the mean absolute logarithmic ratio of Eq.~\eqref{eq:mean_log_difference}. FGPA misses by 17.1\% and 24.9\%. Since FGPA is tuned to the correct mean flux, the gap is not a mean-flux effect. We restrict both suites to half their formal Nyquist wavenumber to reduce sensitivity to discretisation and mass-assignment effects. The retained limits are $k=5.89\,\mathrm{Mpc}^{-1}$ in TNG and $k=2.83\,\mathrm{Mpc}^{-1}$ in FLAMINGO, with $k$ expressed in comoving $\mathrm{Mpc}^{-1}$. Over these ranges, the network errors are 3.6--4.5\%, while FGPA reaches about 35\% in both. The largest relative deviations sit in the highest-$k$ bins, where $P_{\rm 1D}$ falls steeply and small absolute errors become large fractional ones.

We compare the 2LPT and $N$-body inputs over the 13 TNG bins retained below half the Nyquist wavenumber. Their median power spectra differ by $3.4\%$ on average, compared with a characteristic combined checkpoint and paired-patch uncertainty of approximately $4.9\%$. The observed difference is therefore not statistically significant. At this resolution, the cheaper 2LPT input is sufficient for reproducing $P_{\rm 1D}$ through half the Nyquist wavenumber. This conclusion applies to the power spectrum rather than exact field reconstruction, for which the $N$-body model has the smaller flux RMSE.

The most important test is on scales longer than the training box. There are two TNG bins below $2\pi/(110\,\mathrm{Mpc})=0.0571\,\mathrm{Mpc}^{-1}$ and one FLAMINGO bin below $2\pi/(1\,\mathrm{Gpc})=0.00628\,\mathrm{Mpc}^{-1}$; these ranges are marked as colour-coded shaded regions in \cref{fig:pdf_p1d}. The TNG predictions remain accurate at the few-percent level and the FLAMINGO difference is below $1\%$. Direct tracing of the network architectures gives line-of-sight receptive fields of 741 TNG grid cells and 205 FLAMINGO grid cells, corresponding to approximately 180 and $114\,\mathrm{Mpc}$ at the test resolutions. The TNG receptive field wraps across the periodic 448-pixel training skewer, whereas the FLAMINGO receptive field is much smaller than its $1\,\mathrm{Gpc}$ training volume. Long-mode agreement therefore cannot generally be attributed to the receptive field spanning the mode; it is consistent with a local, translation-equivariant mapping applied to input fields that already contain the long-wavelength modulation. Because only two TNG bins and one FLAMINGO bin lie in these ranges, we treat this as an encouraging consistency check rather than a definitive demonstration of long-wavelength extrapolation.

The PDFs agree over most of the flux range. The exception is the continuum, $F\simeq1$, where the FLAMINGO network underpredicts the large population of nearly transparent pixels. It still does markedly better than FGPA. We trace the origin of this deficit in \cref{subsec:discussion_nn_model}. For the same 160-bin, mean-flux-matched PDFs, the Kolmogorov--Smirnov distances are 0.015 and 0.021 for the TNG $N$-body and 2LPT networks, compared with 0.087 for TNG FGPA. In FLAMINGO they are 0.055 for ML and 0.107 for FGPA. The corresponding total-variation distances are 0.026, 0.038, 0.108, 0.084 and 0.135, respectively.

\subsection{Power-spectrum correlation matrices}
\label{subsec:corr}

Correlation matrices test how different $k$-bins of $P_{\rm 1D}$ are coupled. They do not test the variance itself, which normalises them away. Each matrix is estimated from patch-level power spectra averaged over the 64 skewers in each independent $8\times8$ transverse tile, giving 5000 TNG and 1500 FLAMINGO realizations.

\begin{figure*}[!htbp]
\centering
\includegraphics[width=\textwidth]{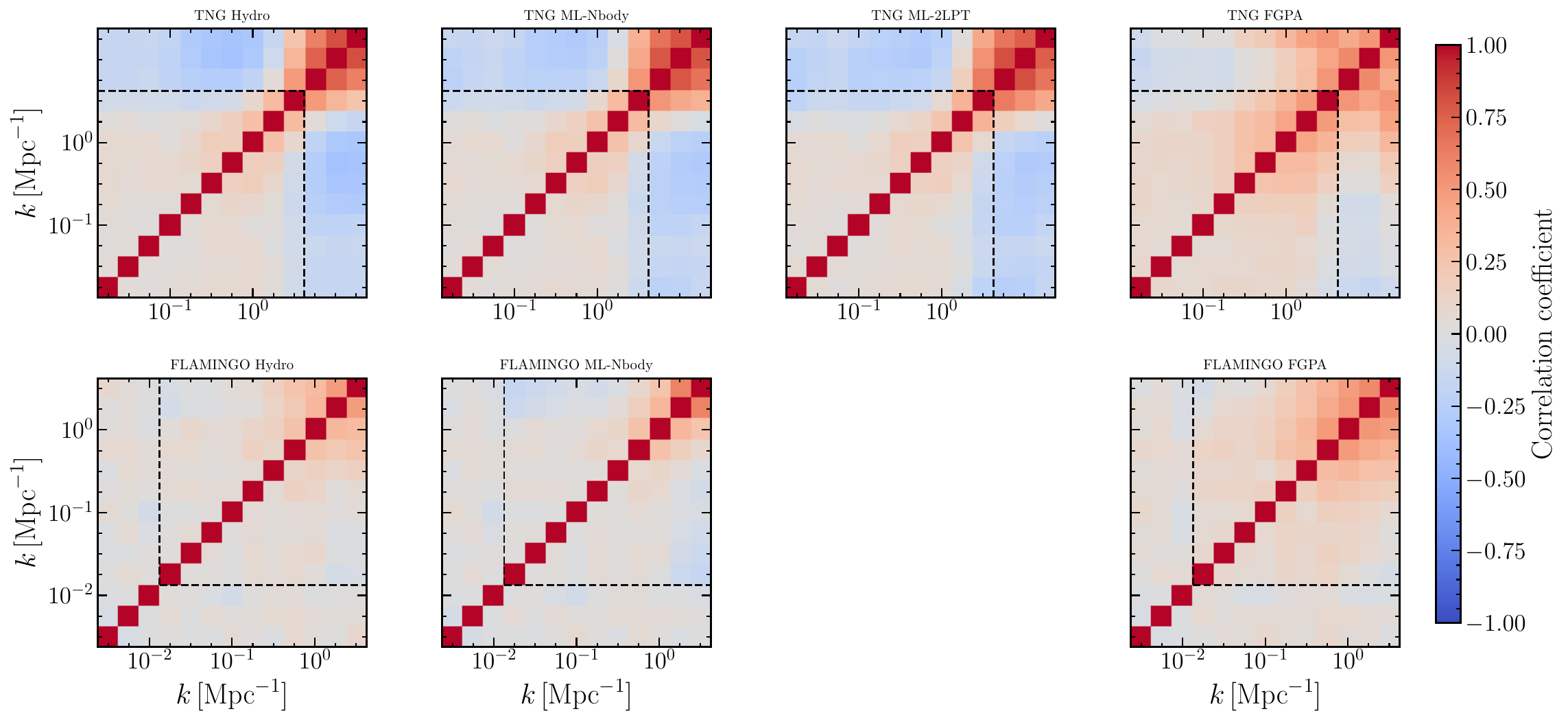}
\caption{Correlation matrices of the noiseless $P_{\rm 1D}$. Top: TNG hydrodynamical, ML ($N$-body and 2LPT inputs) and FGPA. Bottom: FLAMINGO hydrodynamical, ML ($N$-body) and FGPA. Dashed brackets mark the $k$-bins common to both suites.}
\label{fig:both_corr}
\end{figure*}

In TNG the $N$-body and 2LPT networks reproduce the hydrodynamical matrix with mean absolute off-diagonal differences of 0.006 and 0.021 over the common bins (\cref{fig:both_corr}). FLAMINGO recovers the same broad pattern with more element-wise scatter, giving 0.042. The corresponding FGPA--hydro differences are 0.086 in TNG and 0.089 in FLAMINGO. FGPA is therefore substantially less accurate than ML in TNG. In FLAMINGO, ML also improves the element-wise agreement, although the difference between ML and FGPA is smaller and the correlation matrix is consequently a weaker discriminator between methods.

\subsection{Bispectrum and trispectrum}
\label{subsec:higher_order}

\begin{figure}[!htbp]
\centering
\includegraphics[width=\textwidth]{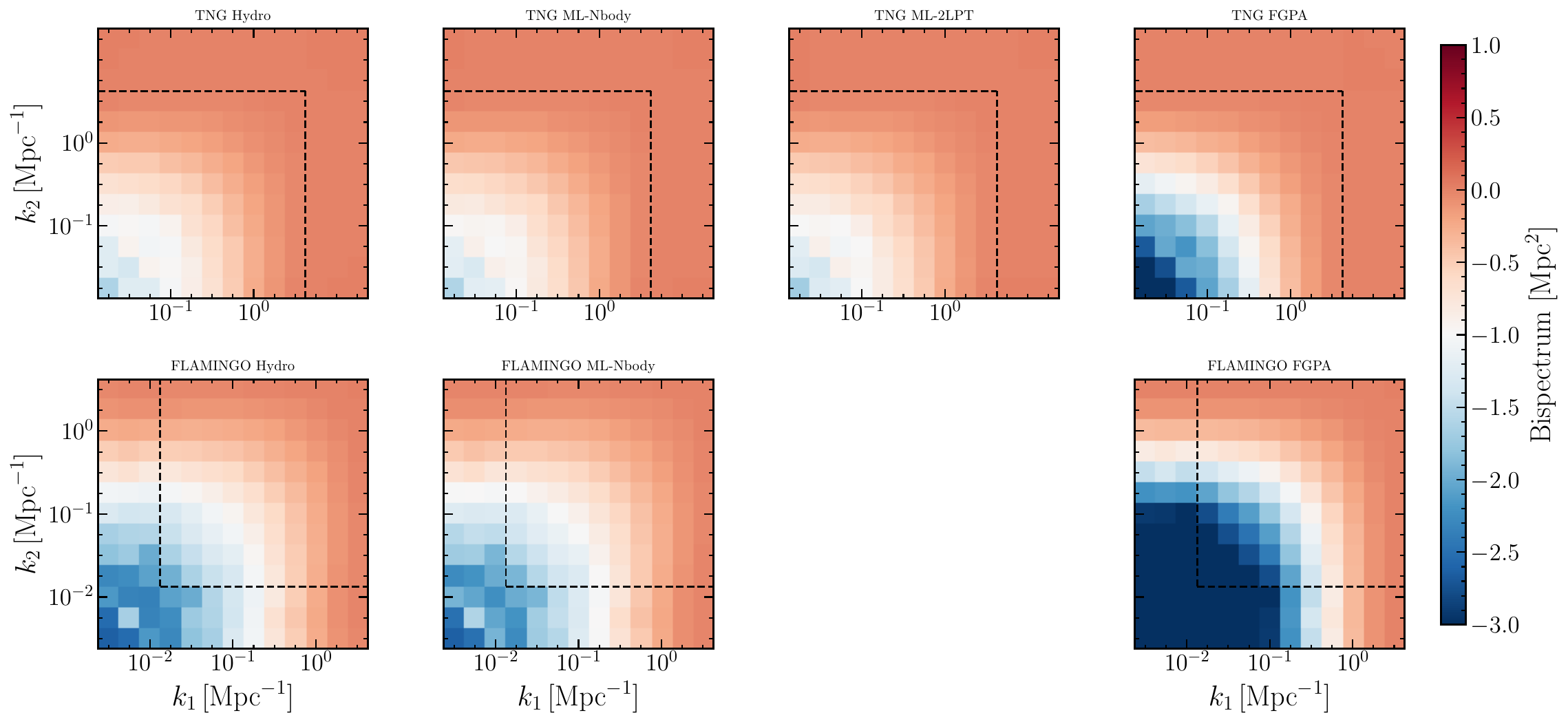}
\caption{1D flux bispectra, binned by mode, in the TNG (top) and FLAMINGO (bottom) test volumes, for the hydrodynamical reference, ML predictions and FGPA. The horizontal and vertical axes are the positive mode-bin centres $k_1$ and $k_2$ in $\mathrm{Mpc}^{-1}$. The colour shows $B(k_1,k_2,-k_1-k_2)$ in $\mathrm{Mpc}^{2}$, averaged over all valid Fourier-mode pairs in each $(k_1,k_2)$ bin satisfying one-dimensional triangle closure.}
\label{fig:bispectrum}
\end{figure}

\begin{figure}[!htbp]
\centering
\includegraphics[width=\textwidth]{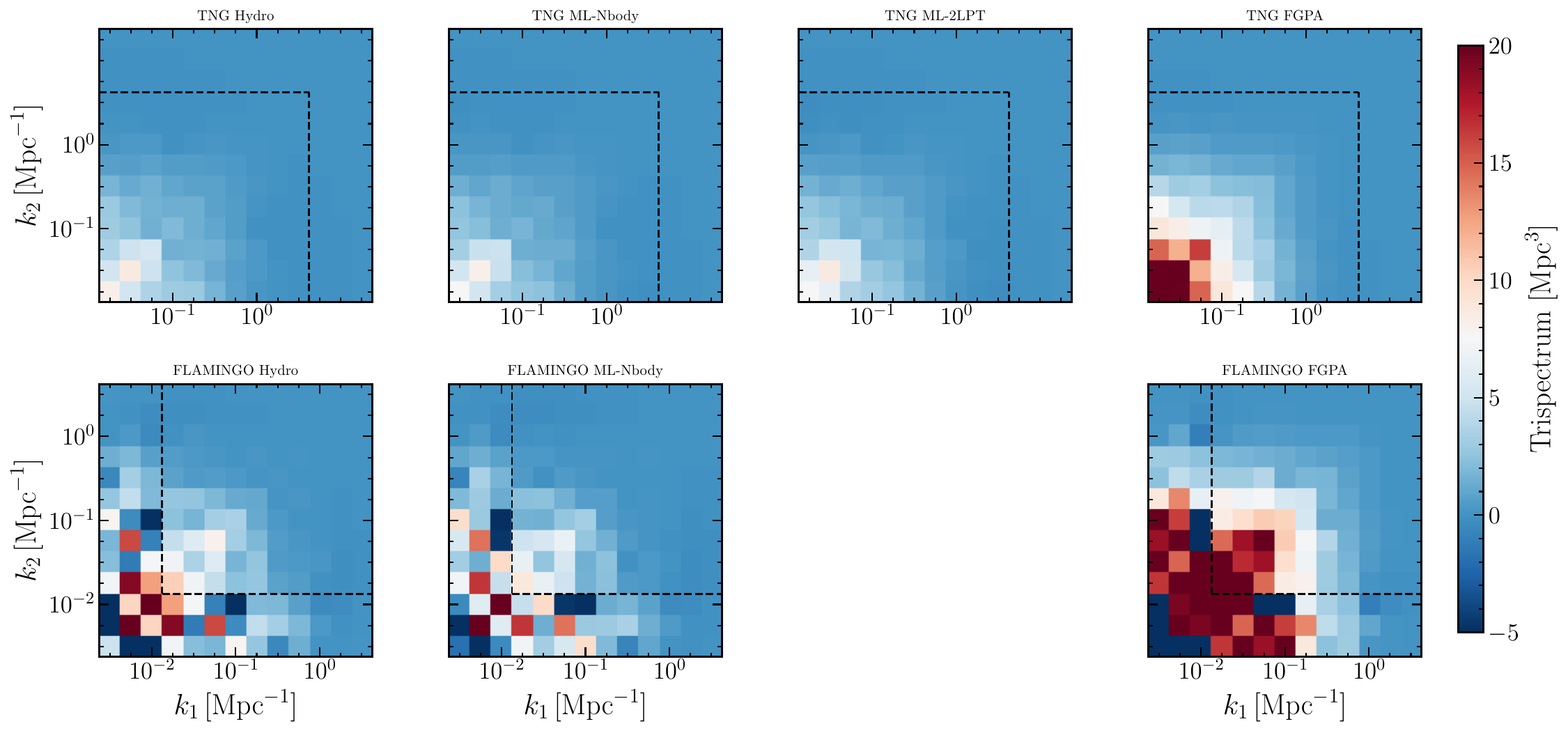}
\caption{Connected 1D flux trispectra, binned by mode; panels as in \cref{fig:bispectrum}. The axes show the positive mode-bin centres $k_1$ and $k_2$ in $\mathrm{Mpc}^{-1}$, and the colour shows the collapsed connected configuration $T_{\rm c}(k_1,-k_1,k_2,-k_2)$ in $\mathrm{Mpc}^{3}$. Each element is the connected covariance of the mode-binned powers, averaged over all positive Fourier modes in the two bins; disconnected Wick contractions, including the additional diagonal exchange term, are subtracted.}
\label{fig:trispectrum}
\end{figure}

These statistics carry the strongest evidence in this paper, because neither enters the loss. \Cref{fig:bispectrum,fig:trispectrum} show both, on a shared colour scale. We compare them over ten common $k$-bins using the symmetric normalised RMS difference of Eq.~\eqref{eq:normalized_rms_difference}, evaluated over the full 2D map. This metric stays well behaved where the signed spectra cross zero. 

The bispectrum is recovered to 5.3\% and 3.9\% in TNG ($N$-body, 2LPT) and 4.4\% in FLAMINGO. FGPA is off by 62--65\%. Hydro-BAM \citep{Sinigaglia2022} also reproduces the three-dimensional Ly$\alpha$ flux bispectrum for the configurations considered up to approximately $0.8\,h\,\mathrm{Mpc}^{-1}$. This provides useful qualitative context but not a quantitative comparison: Hydro-BAM measures a three-dimensional bispectrum parameterized by two side lengths and their enclosed angle, whereas we use the one-dimensional collinear configuration $B(k_1,k_2,-k_1-k_2)$ and quote a normalized RMS difference over the complete common mode-binned map.

The connected trispectrum is harder. The errors grow to 10.0\% and 11.2\% in TNG and 16.0\% in FLAMINGO, against 92--97\% for FGPA, which retains almost no connected four-point signal. To estimate the sampling floor, we randomly divide each hydrodynamical sample into two equal subsets and recompute the complete mode-binned map in each half. Across 20 splits, the median connected-trispectrum differences are $46.3\%$ in TNG and $68.8\%$ in FLAMINGO, with 16th--84th percentile ranges of $34.7$--$60.4\%$ and $56.5$--$83.1\%$. Assuming leading sampling errors scale as $N^{-1/2}$ gives a combined full-sample TNG--FLAMINGO floor of $41.5\%$. The measured $39.8\%$ hydro--hydro difference is therefore noise-limited and should not be interpreted as a resolved physical difference between the simulations.

A useful yardstick is the difference between the two hydrodynamical suites themselves. Over the common bins, TNG and FLAMINGO differ by 15.1\% in $P_{\rm 1D}$, 17.4\% in the bispectrum and 39.8\% in the trispectrum. The bispectrum split-half analysis gives a combined full-sample sampling floor of $4.9\%$, so its $17.4\%$ cross-simulation difference is resolved at the precision of this test. The trispectrum difference, by contrast, is comparable to its $41.5\%$ floor. The emulator errors are smaller than the corresponding raw cross-simulation differences, but only the better-determined statistics support a physical comparison between suites.

\subsection{Summary and comparison with previous work}
\label{subsec:results_summary}

\Cref{tab:results_summary} collects all metrics. The FLAMINGO common-bin $P_{\rm 1D}$ difference exceeds its full-range value because $D_{\rm log}$ weights every bin equally. The full 13-bin range includes three additional large-scale bins below the TNG fundamental mode whose mean difference is only $1.0\%$. The ten-bin common comparison excludes these well-recovered modes while retaining the two highest-$k$ bins, whose differences are $11.9\%$ and $22.3\%$, increasing the unweighted mean from $4.5\%$ to $5.6\%$.

\begin{table*}
\centering
\caption{Emulator and FGPA errors relative to the hydrodynamical
reference. For \(P_{\rm 1D}\), we report the mean absolute logarithmic ratio; for the correlation matrix, the mean absolute off-diagonal difference; and for \(B\) and \(T\), the symmetric normalised RMS difference over the common bins. The final column compares the two hydrodynamical suites.}
\label{tab:results_summary}

\small
\setlength{\tabcolsep}{4pt}
\renewcommand{\arraystretch}{1.25}

\begin{tabularx}{\textwidth}{
    >{\raggedright\arraybackslash}X
    *{3}{>{\centering\arraybackslash}p{0.10\textwidth}}
    |
    *{2}{>{\centering\arraybackslash}p{0.10\textwidth}}
    |
    >{\centering\arraybackslash}p{0.11\textwidth}
}
\hline

\multicolumn{1}{c}{}
& \multicolumn{3}{c}{\textbf{ML vs. Hydro}}
& \multicolumn{2}{c}{\textbf{FGPA vs. Hydro}}
& \multicolumn{1}{c}{\textbf{Hydro}} \\[-2pt]

\cline{2-4}\cline{5-6}
\noalign{\vskip 3pt}

\multicolumn{1}{l}{\textbf{Statistic}}
& \multicolumn{1}{c}{\shortstack{\textbf{TNG}\\\textbf{\(N\)-body}}}
& \multicolumn{1}{c}{\shortstack{\textbf{TNG}\\\textbf{2LPT}}}
& \multicolumn{1}{c}{\shortstack{\textbf{FLAMINGO}\\\textbf{\(N\)-body}}}
& \multicolumn{1}{c}{\textbf{TNG}}
& \multicolumn{1}{c}{\textbf{FLAMINGO}}
& \multicolumn{1}{c}{\shortstack{\textbf{TNG vs.}\\\textbf{FLAMINGO}}} \\

\hline

\(P_{\rm 1D}\), \(k\leq1\,\mathrm{Mpc}^{-1}\)
& 2.7\% & 2.1\% & 2.2\% & 17.1\% & 24.9\% & -- \\

\(P_{\rm 1D}\), full range
& 3.6\% & 4.5\% & 4.5\% & 35.5\% & 34.6\% & -- \\

\(P_{\rm 1D}\), common bins
& 2.8\% & 2.8\% & 5.6\%
& 19.1\% & 32.9\% & 15.1\% \\

Correlation matrix
& 0.006 & 0.021 & 0.042
& 0.086 & 0.089 & 0.055 \\

Bispectrum
& 5.3\% & 3.9\% & 4.4\% & 61.9\% & 64.7\% & 17.4\% \\

Trispectrum
& 10.0\% & 11.2\% & 16.0\% & 92.1\% & 97.1\% & 39.8\% \\

Full-volume inference
& \(\sim 5\,\mathrm{GPU}\)-minutes
& \(\sim 5\,\mathrm{GPU}\)-minutes
& \(\sim 3\,\mathrm{GPU}\)-hours
& -- & -- & -- \\

Hydro cost (million CPU-hours)
& \multicolumn{2}{c}{$\sim$ 35}
& $\sim$ 31
& -- & -- & -- \\

\hline
\end{tabularx}
\end{table*}

We provide a qualitative comparison to previous work as a direct comparison with earlier emulators is difficult, because redshift, resolution and $k$-range all differ. \cite{Jacobus2025} enhanced a low-resolution Gpc-scale hydrodynamical run and found $P_{\rm 1D}$ agreement at the 10--20\% level. LyAl-Net recovers the mean transmitted flux to within 2.5\% \citep{Boonkongkird2023}. \cite{Harrington:2021srm} reported approximately $5\%$ power-spectrum agreement to $k\simeq10\,h\,\mathrm{Mpc}^{-1}$, while \cite{Peirani2022} achieved approximately $2\%$ to $k\simeq4\,h\,\mathrm{Mpc}^{-1}$ after iterative corrections. These values are not directly comparable with ours because the studies use different error definitions, resolutions, $k$ ranges and calibration procedures. Earlier fully convolutional models could in principle be deployed on larger boxes, but our analysis directly validates the prediction against independent hydrodynamical volumes approximately twenty times larger than the training volumes and additionally tests the connected trispectrum.


\section{Discussion}
\label{sec:discussion}

\subsection{The optical-depth tails}
\label{subsec:discussion_nn_model}

The one visible failure in the results is the flux PDF near the continuum in FLAMINGO. It is easier to understand in $\log\tau$ than in $F$. Because $F=e^{-\tau}$ compresses all low-$\tau$ pixels towards $F=1$, a modest deficit of very transparent pixels appears as a pronounced dip at the continuum.

\begin{figure}
\centering
\includegraphics[width=0.8\textwidth]{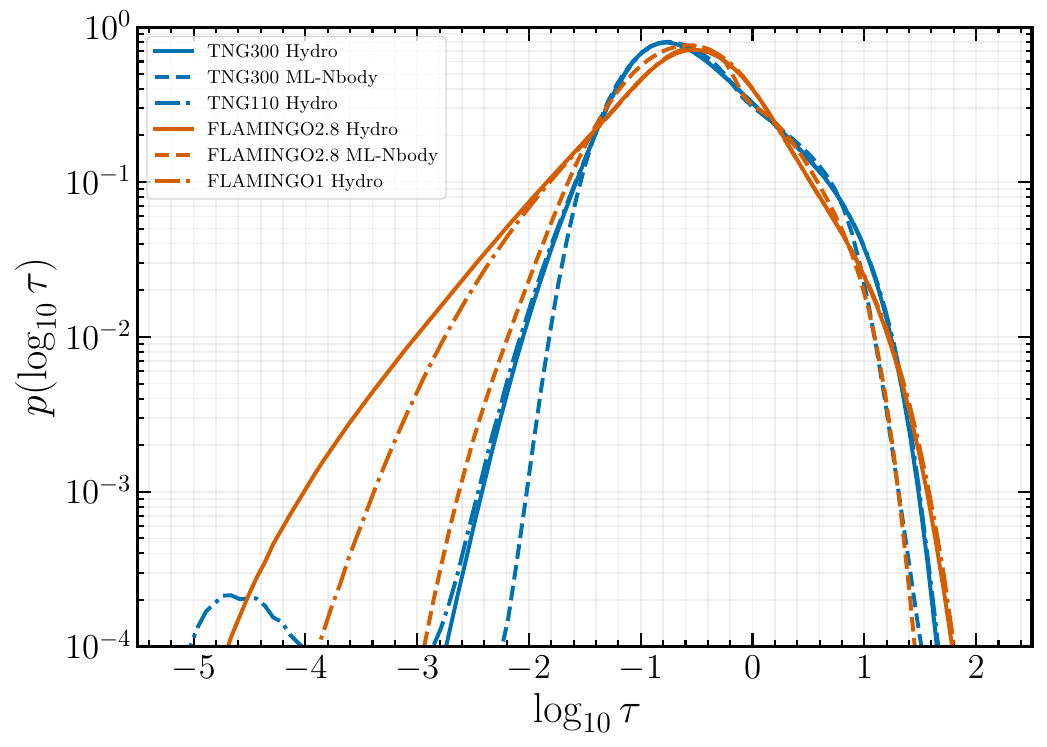}
\caption{Distributions of $\log\tau$ in the TNG and FLAMINGO training and test volumes, with ML predictions for the test volumes. The TNG distributions are converged between 110 and $300\,\mathrm{cMpc}$. The FLAMINGO $2.8\,\mathrm{cGpc}$ box has a longer low-$\tau$ tail than the $1\,\mathrm{cGpc}$ box, which the network underpredicts, producing the deficit near $F\simeq1$.}
\label{fig:logtau_pdf}
\end{figure}

\Cref{fig:logtau_pdf} shows the $\log\tau$ distributions. In TNG the 110 and $300\,\mathrm{cMpc}$ boxes have nearly identical distributions, and the network reproduces the larger one apart from a slight underprediction in both tails. FLAMINGO behaves differently. The $2.8\,\mathrm{cGpc}$ box has a noticeably longer low-$\tau$ tail than the $1\,\mathrm{cGpc}$ training box, and this is the part the network misses.

The FLAMINGO network is trained on \(400\times8\times8\times1800\simeq4.6\times10^7\) voxels, substantially fewer than the number of voxels in the complete \(1\,\mathrm{cGpc}\) volume. The sampled training set may therefore under-represent rare void environments even when they are present in the parent box. The \(1\,\mathrm{cGpc}\) and \(2.8\,\mathrm{cGpc}\) FLAMINGO simulations use identical calibrations, subgrid modelling, and ultraviolet-background prescriptions, so the difference in their low-\(\tau\) tails cannot be attributed to changes in these ingredients. Limited sampling of the low-density tail therefore provides a direct explanation for why the much larger test volume exposes low-\(\tau\) configurations that the network reproduces less accurately.

There is also a generic cause. A network trained with a mean-squared error tends toward the conditional mean of $\log\tau$ given the inputs. This averaging narrows the predicted distribution and can suppress both tails, consistent with the milder effect seen in TNG. The power-spectrum term partly counteracts this behaviour but does not explicitly constrain one-point extremes. Possible remedies include adding a PDF- or quantile-based loss term, oversampling rare environments, or adopting a stochastic output model.

\subsection{Implications for mock catalogues}
\label{subsec:discussion_mocks}

Survey analyses need large volumes, many realisations and correct correlations between scales. Hydrodynamical simulations deliver the physics but not the volume. Analytic prescriptions deliver the volume but, as the FGPA comparison shows, lose most of the non-Gaussian structure: the trispectrum, which sets the non-Gaussian covariance, is essentially absent from FGPA fields. The emulator sits between the two. It runs on gravity-only fields, which are cheap enough for many Gpc-scale realisations, and keeps the hydrodynamical statistics to within a few per cent in the bispectrum and about 10--16\% in the trispectrum. Extrapolating our measured throughput to non-overlapping $8\times8$ patches over the complete transverse grids gives approximately $\sim 5$ GPU-minutes for TNG300 and $\sim 3$ GPU-hours for FLAMINGO2.8, including a modest allowance for data loading and output. The corresponding hydrodynamical simulations required approximately 34.9 and 31 million CPU core-hours, respectively. These are not like-for-like costs: hydrodynamics evolves the complete physical system and produces many outputs, whereas inference requires precomputed density and velocity fields. The comparison nevertheless demonstrates the small incremental cost of generating a full-volume flux field once the inputs and trained network are available.

Two applications stand out. The first is covariance estimation for $P_{\rm 1D}$ and 3D forest analyses, where non-Gaussian mode coupling matters and many realisations are needed. The second is small-scale cosmology. Constraints on warm dark matter or massive neutrinos depend on the shape of $P_{\rm 1D}$ at high $k$ and on how modes are correlated, and a method that distorts either can bias the result.The present networks are trained at a single redshift, cosmology and thermal history per suite and therefore cannot yet be used directly for parameter inference in such models. That application will require training or conditioning across the relevant cosmological and astrophysical parameter space.

\subsection{Limitations}
\label{subsec:limitations}

\paragraph{Transfer between simulation suites.}
Cross-simulation transfer is more demanding than volume transfer within a single simulation suite. To remove the direct mismatch in spatial resolution, we downsample the TNG300 density, velocity, and optical-depth fields to the FLAMINGO grid resolution and train a new network on these downsampled TNG300 fields. We then apply this resolution-matched TNG300 network to the \(1\,\mathrm{cGpc}\) FLAMINGO volume without any further retraining. Finally, we rescale the predicted optical depths so that the resulting mean transmitted flux matches the FLAMINGO hydrodynamical value.

\begin{figure}
\centering
\includegraphics[width=\textwidth]{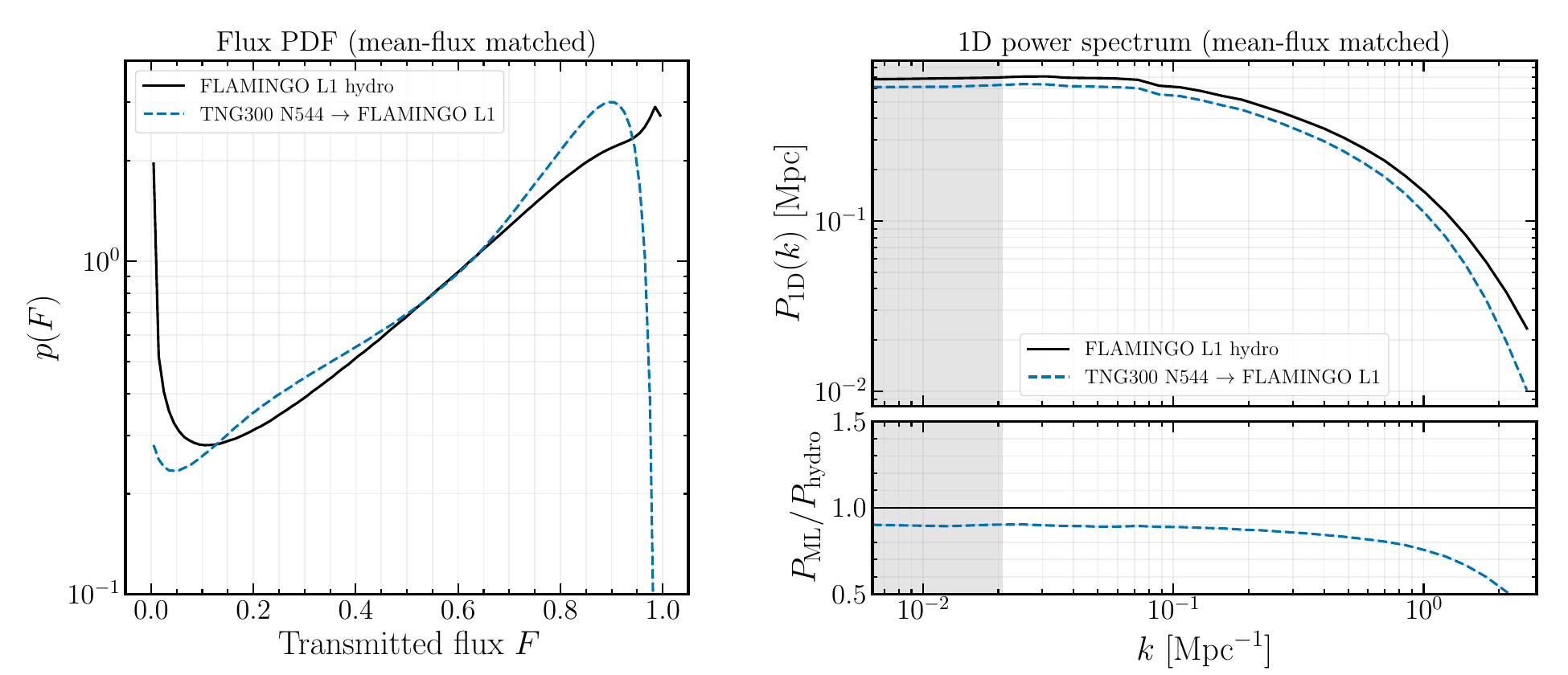}
\caption{Mean-flux-matched flux statistics in the FLAMINGO \(1\,\mathrm{cGpc}\) box. The solid black curves show the FLAMINGO hydrodynamical reference, while the dashed blue curves show the prediction from the network trained on the \(N544\) TNG300 simulation. The left panel compares the flux PDFs: the prediction agrees reasonably well over the intermediate-flux range but over-predicts the distribution around \(F\simeq0.8\)–\(0.95\) and under-predicts pixels close to the continuum. The upper-right panel shows \(P_{\rm 1D}\), with the prediction-to-hydrodynamical ratio displayed below. The predicted power is within \(\sim 10\%\) on scales larger than training volume but becomes progressively more suppressed toward high \(k\). The shaded region marks \(k<2\pi/(300\,\mathrm{cMpc})\), corresponding to wavelengths larger than the TNG300 training box.}
\label{fig:cross_sim_pdf_p1d}
\end{figure}

The agreement is poorer than in the within-suite volume-transfer tests, but the network still recovers the large-scale \(P_{\rm 1D}\) to approximately \(10\%\); see \cref{fig:cross_sim_pdf_p1d}. This includes the three modes with \(k<2\pi/L_{\rm train}\), corresponding to wavelengths larger than the \(302.6\,\mathrm{cMpc}\) TNG300 training volume. The agreement therefore indicates that the network retains useful large-scale predictive power across simulation suites, although it should not be interpreted as precise validation from only three long-wavelength bins. At progressively smaller scales, the prediction increasingly underestimates the FLAMINGO power, showing that matching the grid resolution and mean flux does not fully remove the domain shift between the simulations.

The mean-flux-matched flux PDF also reproduces the broad shape of the FLAMINGO distribution, particularly at intermediate transmitted flux, but differs near the low- and high-flux tails and around the peak close to \(F\simeq0.9\). Quantitatively, the binned Kolmogorov--Smirnov distance,
\[
D_{\rm KS}=\max_F\left|C_{\rm ML}(F)-C_{\rm hydro}(F)\right|,
\]
is \(D_{\rm KS}\simeq0.10\), where \(C(F)\) denotes the cumulative flux distribution. Thus, mean-flux matching fixes the first moment of the distribution but does not guarantee agreement in its complete shape.

\begin{figure}
\centering
\includegraphics[width=\textwidth]{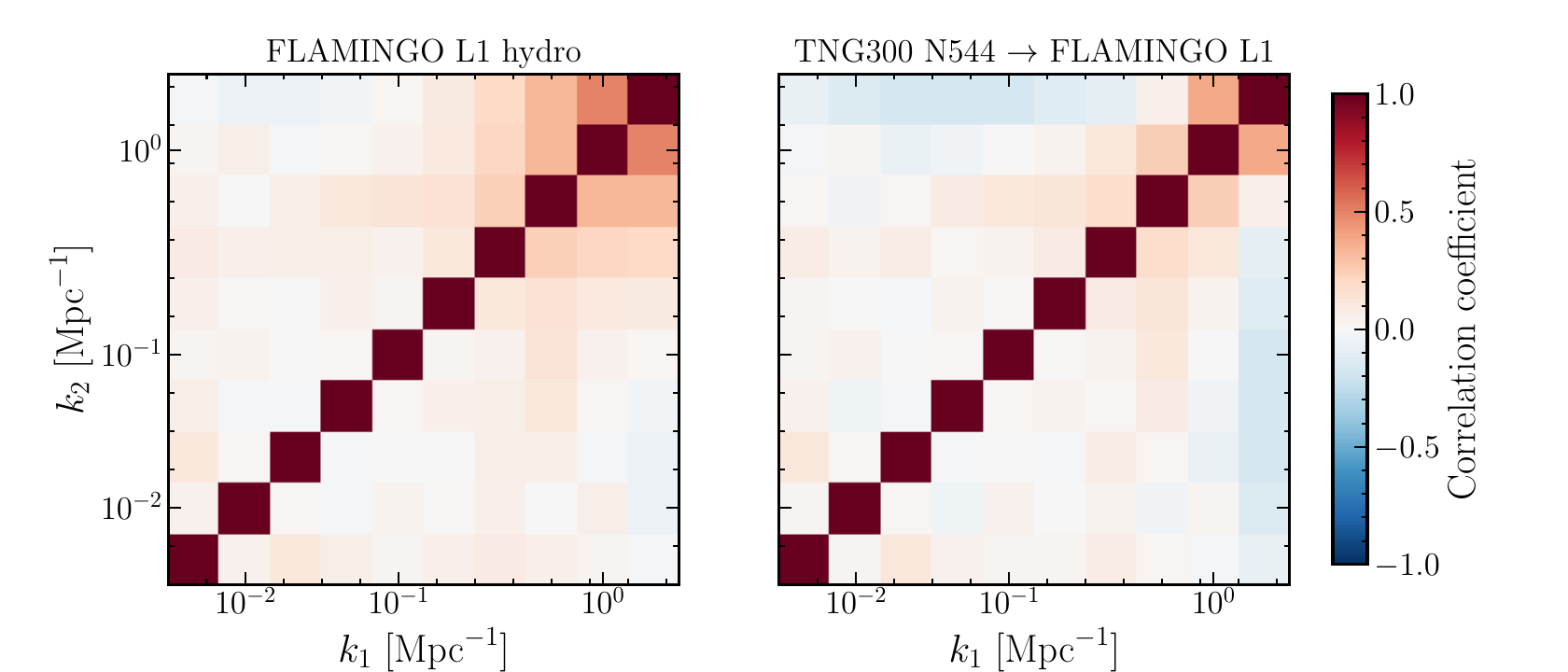}
\caption{Correlation matrices of the \(P_{\rm 1D}\) in the FLAMINGO \(1\,\mathrm{cGpc}\) box. The left panel shows the FLAMINGO hydrodynamical reference, while the right panel shows the prediction from the network trained on the \(N544\) TNG300 simulation. The cross-simulation prediction reproduces the weak correlations across much of the large-scale region but does not recover the hydrodynamical mode-coupling pattern accurately at smaller scales. In particular, it under-predicts the positive correlations involving the highest-\(k\) bins and introduces weak negative correlations between these bins and larger-scale modes.}
\label{fig:cross_sim_corr}
\end{figure}

The \(P_{\rm 1D}\) correlation matrices show a similar result (fig.~\ref{fig:cross_sim_corr}). The transferred model recovers the dominant diagonal structure and the general increase in positive correlations among the highest-\(k\) bins. The mean absolute difference between the off-diagonal correlation coefficients is
\[
\left\langle
\left|r_{ij}^{\rm ML}-r_{ij}^{\rm hydro}\right|
\right\rangle_{i\ne j}=0.056.
\]
Nevertheless, visible differences remain in correlations involving the highest-wavenumber bins. The model therefore captures the broad covariance structure across simulations, but not all of its detailed scale dependence.

Because the grid resolution and mean transmitted flux have been explicitly matched, the remaining discrepancies cannot be attributed primarily to those two effects. They instead indicate sensitivity to differences between the TNG and FLAMINGO domains, including their cosmologies, thermal and ionization histories, feedback prescriptions, numerical implementations, and the resulting mapping from density and velocity to optical depth. Cross-suite performance could potentially be improved by training on multiple hydrodynamical simulations at a common resolution, conditioning the model on thermal and ionization parameters, or applying domain-adaptation methods.

\paragraph{Coverage of environments.}
Recovering modes longer than the training box shows that the network is not memorising Fourier modes. It does not show that the network can extrapolate to environments it rarely saw. A training box, or a training sample drawn from it, may under-represent the deepest voids, the densest filaments, or unusual velocity configurations. The network then learns a smoothed mapping in those regions, and a much larger test box exposes the gap. The FLAMINGO low-$\tau$ tail is likely an example. Larger or more diverse training sets, sampling that deliberately oversamples rare regions, and loss terms that constrain the one-point distribution are all natural remedies.

\paragraph{Scope.}
All results are at \(z\sim3\), for a single cosmology and thermal history per suite, and are based on noiseless skewers. The grid spacings, \(0.24\,\mathrm{cMpc}\) in TNG and \(0.56\,\mathrm{cMpc}\) in FLAMINGO, are substantially coarser than the \(20\,h^{-1}\,\mathrm{ckpc}\) resolution found to be necessary for approximately \(1-2\%\) convergence of Ly\(\alpha\)-forest flux statistics up to \(k \simeq 10\,h\,\mathrm{Mpc}^{-1}\) in dedicated convergence studies \citep{Lukic2015}. They are also coarser than the \(\sim20\,\mathrm{kpc}\) resolution adopted for high-fidelity Ly\(\alpha\)-forest reconstruction by \cite{Harrington:2021srm}. Our results should therefore be interpreted as applying to the scales resolved by the grids used here, rather than as predictions of converged high-\(k\) forest statistics. Moreover, the present analysis does not include continuum-fitting uncertainties, observational noise, or finite spectrograph resolution. A comparison with observational data would also require identifying and masking or modelling damped Ly\(\alpha\) absorbers and metal-line contamination. These effects must be incorporated through realistic forward modelling before the emulator can be tested robustly against observations. Accounting for them will be an important direction for future work.


\section{Conclusion}
\label{sec:conclusion}

We have trained a three-dimensional U-Net to predict the Ly\(\alpha\) optical depth from gravity-only density and line-of-sight velocity fields. Our primary aim was to determine whether a mapping calibrated on a relatively small hydrodynamical simulation remains valid when applied to a substantially larger volume. We tested this in two simulation suites: from \(110\) to \(300\,\mathrm{cMpc}\) in IllustrisTNG, using both 2LPT and \(N\)-body inputs, and from \(1\) to \(2.8\,\mathrm{cGpc}\) in FLAMINGO, using \(N\)-body inputs. In both cases, the test volume is approximately twenty times larger than the corresponding training volume.

Within a given simulation suite, the network transfers successfully to the larger volume. Our main findings are as follows:
\begin{itemize}
    \item The 1D flux power spectrum is recovered to \(2\)--\(3\%\) for \(k\leq1\,\mathrm{Mpc}^{-1}\) and to better than \(5\%\) over the full retained range, compared with discrepancies of \(17\)--\(35\%\) for FGPA. Modes with wavelengths longer than the training box are recovered with comparable accuracy.
    
    \item The bispectrum, which is not included explicitly in the training objective, is recovered to \(4\)--\(5\%\), while the connected trispectrum is recovered to \(10\)--\(16\%\). The corresponding FGPA discrepancies are approximately \(60\)--\(97\%\).
    
    \item For every statistic considered, the emulator error within each suite is smaller than the difference between the TNG and FLAMINGO hydrodynamical results. The networks therefore recover statistical features specific to their respective hydrodynamical training suites rather than merely generating a common average Ly\(\alpha\)-forest field.
    
    \item The 2LPT- and \(N\)-body-based TNG networks yield statistically consistent \(P_{\rm 1D}\) predictions up to half the TNG Nyquist wavenumber. Their mean difference is \(3.4\%\), below the approximate \(4.3\%\) combined uncertainty floor, indicating that the less expensive 2LPT inputs are sufficient at the resolution considered here.
\end{itemize}

Two limitations remain. First, the network under-predicts the low-\(\tau\) tail in the largest FLAMINGO volume, producing a deficit of highly transparent pixels near \(F\simeq1\). The likely causes are incomplete sampling of rare low-density environments and the tendency of an MSE-trained deterministic model to predict a conditional mean, thereby suppressing the extremes of the output distribution.

Second, transfer between different simulation suites remains substantially more difficult than volume transfer within a single suite. To control for spatial resolution, we train a network on TNG300 fields downsampled to the FLAMINGO grid resolution and apply it to the \(1\,\mathrm{cGpc}\) FLAMINGO volume. Even after matching the mean transmitted flux, the prediction underestimates \(P_{\rm 1D}\) by approximately \(10\%\) on large scales, with the suppression increasing toward smaller scales. The broad power spectrum correlation structure is reproduced reasonably well on large scales, but the detailed small-scale mode coupling is not: the prediction underestimates positive correlations involving the highest-\(k\) bins and introduces weak anti-correlations with larger-scale modes. Because the grid resolutions are matched in this test, the remaining discrepancies point to differences between the suites in cosmology and in the physical mapping from density and velocity to Ly\(\alpha\) absorption, including their thermal histories, feedback prescriptions, and hydrodynamical modelling.

Cheap gravity-only simulations can therefore be turned into Gpc-scale Ly$\alpha$ forest mocks that keep most of the hydrodynamical statistical structure, including the non-Gaussian mode coupling relevant to covariance estimation for DESI and future surveys. The next steps are clear: conditioning on cosmological and thermal parameters, training across simulation suites, and constraining the one-point distribution directly.

\section*{Acknowledgements}
The computations presented in this work were carried out using the Param Sanganak high-performance computing facility at the Indian Institute of Technology Kanpur and the Pawna high-performance computing facility at the Tata Institute of Fundamental Research, Mumbai. The authors appreciate the support of these computing resources. This work was supported by the Department of Atomic Energy, Government of India, under Project Identification Number RTI-4012. We also thank Kapil Ghadiali and Ajay Salve for their support with the HPC facility at TIFR, Mumbai.

We acknowledge the use of AI assistants (OpenAI's ChatGPT, Google Gemini and Anthropic's Claude) for help with writing and code assistance during the preparation of this manuscript. All scientific content, analysis, and conclusions are the authors' own and have been verified by the authors.

\section*{Data Availability}
The data generated during this work will be made available upon reasonable request to the authors.

\bibliographystyle{JHEP}
\bibliography{references}

@article{Rauch1998,
  author  = {Rauch, Michael},
  title   = {The Lyman Alpha Forest in the Spectra of Quasistellar Objects},
  journal = {Annual Review of Astronomy and Astrophysics},
  volume  = {36},
  pages   = {267--316},
  year    = {1998},
  doi     = {10.1146/annurev.astro.36.1.267}
}

@article{Weinberg1997,
  author        = {Weinberg, David H. and Hernquist, Lars and Katz, Neal and Croft, Rupert and Miralda-Escud{\'e}, Jordi},
  title         = {Hubble Flow Broadening of the Lyman-alpha Forest and its Implications},
  journal       = {arXiv e-prints},
  pages         = {astro-ph/9709303},
  year          = {1997},
  eprint        = {astro-ph/9709303},
  archivePrefix = {arXiv}
}

@article{Croft2002,
  author        = {Croft, Rupert A. C. and Weinberg, David H. and Bolte, Michael and Burles, Scott and Hernquist, Lars and Katz, Neal and Kirkman, David and Tytler, David},
  title         = {Towards a Precise Measurement of Matter Clustering: Lyman-alpha Forest Data at Redshifts 2--4},
  journal       = {The Astrophysical Journal},
  volume        = {581},
  pages         = {20--52},
  year          = {2002},
  doi           = {10.1086/344099},
  eprint        = {astro-ph/0012324},
  archivePrefix = {arXiv}
}

@article{McDonald2006,
  author        = {McDonald, Patrick and others},
  title         = {The Lyman-alpha Forest Power Spectrum from the Sloan Digital Sky Survey},
  journal       = {The Astrophysical Journal Supplement Series},
  volume        = {163},
  pages         = {80--109},
  year          = {2006},
  doi           = {10.1086/444361},
  eprint        = {astro-ph/0405013},
  archivePrefix = {arXiv}
}

@article{PalanqueDelabrouille2013,
  author        = {Palanque-Delabrouille, Nathalie and others},
  title         = {The one-dimensional Ly-alpha forest power spectrum from BOSS},
  journal       = {Astronomy \& Astrophysics},
  volume        = {559},
  pages         = {A85},
  year          = {2013},
  doi           = {10.1051/0004-6361/201322130},
  eprint        = {1306.5896},
  archivePrefix = {arXiv}
}

@article{Chabanier2019,
  author        = {Chabanier, Sol{\`e}ne and others},
  title         = {The one-dimensional power spectrum from the SDSS DR14 Ly-alpha forests},
  journal       = {Journal of Cosmology and Astroparticle Physics},
  volume        = {2019},
  number        = {07},
  pages         = {017},
  year          = {2019},
  doi           = {10.1088/1475-7516/2019/07/017},
  eprint        = {1812.03554},
  archivePrefix = {arXiv}
}

@article{Walther2019,
  author        = {Walther, Michael and O{\~n}orbe, Jose and Hennawi, Joseph F. and Luki{\'c}, Zarija},
  title         = {New Constraints on IGM Thermal Evolution from the Ly-alpha Forest Power Spectrum},
  journal       = {The Astrophysical Journal},
  volume        = {872},
  pages         = {13},
  year          = {2019},
  doi           = {10.3847/1538-4357/aafad1},
  eprint        = {1808.04367},
  archivePrefix = {arXiv}
}

@article{Bosman2022,
  author        = {Bosman, Sarah E. I. and others},
  title         = {Measurements of the \(z > 5\) Lyman-alpha forest flux auto-correlation functions from the extended XQR-30 data set},
  journal       = {Monthly Notices of the Royal Astronomical Society},
  volume        = {514},
  pages         = {55--76},
  year          = {2022},
  doi           = {10.1093/mnras/stac1046},
  eprint        = {2108.03699},
  archivePrefix = {arXiv}
}

@article{PalanqueDelabrouille2015,
  author        = {Palanque-Delabrouille, Nathalie and Y{\`e}che, Christophe and Baur, Julien and Magneville, Christophe and Rossi, Graziano and Lesgourgues, Julien and Borde, Arnaud and Burtin, Etienne and LeGoff, Jean-Marc and Rich, James and Viel, Matteo and Weinberg, David},
  title         = {Neutrino masses and cosmology with Lyman-alpha forest power spectrum},
  journal       = {Journal of Cosmology and Astroparticle Physics},
  volume        = {2015},
  number        = {11},
  pages         = {011},
  year          = {2015},
  doi           = {10.1088/1475-7516/2015/11/011},
  eprint        = {1506.05976},
  archivePrefix = {arXiv}
}

@article{Viel2013,
  author        = {Viel, Matteo and Becker, George D. and Bolton, James S. and Haehnelt, Martin G.},
  title         = {Warm dark matter as a solution to the small scale crisis: New constraints from high redshift Lyman-alpha forest data},
  journal       = {Physical Review D},
  volume        = {88},
  pages         = {043502},
  year          = {2013},
  doi           = {10.1103/PhysRevD.88.043502},
  eprint        = {1306.2314},
  archivePrefix = {arXiv}
}

@article{Baur2016,
  author        = {Baur, Julien and Palanque-Delabrouille, Nathalie and Y{\`e}che, Christophe and Magneville, Christophe and Viel, Matteo},
  title         = {Lyman-alpha Forests cool Warm Dark Matter},
  journal       = {Journal of Cosmology and Astroparticle Physics},
  volume        = {2016},
  number        = {08},
  pages         = {012},
  year          = {2016},
  doi           = {10.1088/1475-7516/2016/08/012},
  eprint        = {1512.01981},
  archivePrefix = {arXiv}
}

@article{Schaye2000,
  author        = {Schaye, Joop and Theuns, Tom and Rauch, Michael and Efstathiou, George and Sargent, Wallace L. W.},
  title         = {The thermal history of the intergalactic medium},
  journal       = {Monthly Notices of the Royal Astronomical Society},
  volume        = {318},
  pages         = {817--826},
  year          = {2000},
  doi           = {10.1046/j.1365-8711.2000.03815.x},
  eprint        = {astro-ph/9912432},
  archivePrefix = {arXiv}
}

@article{Becker2011,
  author        = {Becker, George D. and Bolton, James S. and Haehnelt, Martin G. and Sargent, Wallace L. W.},
  title         = {Detection of extended He II reionization in the temperature evolution of the intergalactic medium},
  journal       = {Monthly Notices of the Royal Astronomical Society},
  volume        = {410},
  pages         = {1096--1112},
  year          = {2011},
  doi           = {10.1111/j.1365-2966.2010.17507.x},
  eprint        = {1008.2622},
  archivePrefix = {arXiv}
}

@article{Boera2014,
  author        = {Boera, Elisa and Murphy, Michael T. and Becker, George D. and Bolton, James S.},
  title         = {The thermal history of the intergalactic medium down to redshift z = 1.5: a new curvature measurement},
  journal       = {Monthly Notices of the Royal Astronomical Society},
  volume        = {441},
  pages         = {1916--1933},
  year          = {2014},
  doi           = {10.1093/mnras/stu660},
  eprint        = {1404.1083},
  archivePrefix = {arXiv}
}

@article{Worseck2011,
  author        = {Worseck, G{\'a}bor and Prochaska, J. Xavier and McQuinn, Matthew and Dall'Aglio, Aldo and Fechner, Cora and Hennawi, Joseph F. and Reimers, Dieter and Richter, Philipp and Wisotzki, Lutz},
  title         = {The End of Helium Reionization at z ~ 2.7 Inferred from Cosmic Variance in HST/COS He II Lyman-alpha Absorption Spectra},
  journal       = {The Astrophysical Journal Letters},
  volume        = {733},
  pages         = {L24},
  year          = {2011},
  doi           = {10.1088/2041-8205/733/2/L24},
  eprint        = {1103.5752},
  archivePrefix = {arXiv}
}

@article{Worseck2016,
  author        = {Worseck, G{\'a}bor and Davies, Frederick B. and Hennawi, Joseph F. and Prochaska, J. Xavier},
  title         = {The Evolution of the He II-ionizing Background at Redshifts 2.3 < z < 3.8 Inferred from a Statistical Sample of 24 HST/COS He II Lyman-alpha Absorption Spectra},
  journal       = {The Astrophysical Journal},
  volume        = {825},
  pages         = {144},
  year          = {2016},
  doi           = {10.3847/0004-637X/825/2/144},
  eprint        = {1602.04847},
  archivePrefix = {arXiv}
}

@article{Boera2019,
  author        = {Boera, Elisa and Becker, George D. and Bolton, James S. and Nasir, Fahad},
  title         = {Revealing reionization with the thermal history of the intergalactic medium: new constraints from the Ly-alpha flux power spectrum},
  journal       = {The Astrophysical Journal},
  volume        = {872},
  pages         = {101},
  year          = {2019},
  doi           = {10.3847/1538-4357/aafee4},
  eprint        = {1809.06980},
  archivePrefix = {arXiv}
}

@article{Villasenor2021,
  author        = {Villasenor, Bruno and Robertson, Brant and Madau, Piero and Schneider, Evan},
  title         = {Inferring the Thermal History of the Intergalactic Medium from the Properties of the Hydrogen and Helium Lyman-alpha Forest},
  journal       = {The Astrophysical Journal},
  volume        = {933},
  pages         = {59},
  year          = {2022},
  doi           = {10.3847/1538-4357/ac7045},
  eprint        = {2111.00019},
  archivePrefix = {arXiv}
}

@article{York2000,
  author        = {York, Donald G. and others},
  title         = {The Sloan Digital Sky Survey: Technical Summary},
  journal       = {The Astronomical Journal},
  volume        = {120},
  pages         = {1579--1587},
  year          = {2000},
  doi           = {10.1086/301513},
  eprint        = {astro-ph/0006396},
  archivePrefix = {arXiv}
}

@article{Dawson2013,
  author        = {Dawson, Kyle S. and others},
  title         = {The Baryon Oscillation Spectroscopic Survey of SDSS-III},
  journal       = {The Astronomical Journal},
  volume        = {145},
  pages         = {10},
  year          = {2013},
  doi           = {10.1088/0004-6256/145/1/10},
  eprint        = {1208.0022},
  archivePrefix = {arXiv}
}

@article{Dawson2016,
  author        = {Dawson, Kyle S. and others},
  title         = {The SDSS-IV Extended Baryon Oscillation Spectroscopic Survey: Overview and Early Data},
  journal       = {The Astronomical Journal},
  volume        = {151},
  pages         = {44},
  year          = {2016},
  doi           = {10.3847/0004-6256/151/2/44},
  eprint        = {1508.04473},
  archivePrefix = {arXiv}
}

@article{DESI2016,
  author        = {DESI Collaboration and Aghamousa, Amir and others},
  title         = {The DESI Experiment Part I: Science, Targeting, and Survey Design},
  journal       = {arXiv e-prints},
  pages         = {arXiv:1611.00036},
  year          = {2016},
  doi           = {10.48550/arXiv.1611.00036},
  eprint        = {1611.00036},
  archivePrefix = {arXiv}
}

@article{DESI2019,
  author        = {Levi, Michael E. and Allen, Lori E. and Raichoor, Anand and others},
  title         = {The Dark Energy Spectroscopic Instrument \(DESI\)},
  journal       = {Bulletin of the American Astronomical Society},
  volume        = {51},
  number        = {7},
  pages         = {57},
  year          = {2019},
  eprint        = {1907.10688},
  archivePrefix = {arXiv}
}

@article{DESI2023EDR,
  author        = {DESI Collaboration and Adame, A. G. and others},
  title         = {The Early Data Release of the Dark Energy Spectroscopic Instrument},
  journal       = {arXiv e-prints},
  pages         = {arXiv:2306.06308},
  year          = {2023},
  doi           = {10.48550/arXiv.2306.06308},
  eprint        = {2306.06308},
  archivePrefix = {arXiv}
}

@article{Pieri2016,
  author        = {Pieri, Matthew M. and Bonoli, Silvia and Chaves-Montero, Jon{\'a}s and Paris, Isabelle and Fumagalli, Michele and Bolton, James S. and Viel, Matteo and Noterdaeme, Pasquier and Miralda-Escud{\'e}, Jordi and Busca, Nicol{\'a}s G. and Rahmani, Hadi and P{\'e}roux, C{\'e}line and Font-Ribera, Andreu and Trager, Scott C. and The WEAVE Collaboration},
  title         = {WEAVE-QSO: A Massive Intergalactic Medium Survey for the William Herschel Telescope},
  journal       = {arXiv e-prints},
  pages         = {arXiv:1611.09388},
  year          = {2016},
  doi           = {10.48550/arXiv.1611.09388},
  eprint        = {1611.09388},
  archivePrefix = {arXiv}
}

@article{Jin2024,
  author        = {Jin, Shiyin and others},
  title         = {The wide-field, multiplexed, spectroscopic facility WEAVE: Survey design, overview, and simulated implementation},
  journal       = {Monthly Notices of the Royal Astronomical Society},
  volume        = {530},
  pages         = {2688--2736},
  year          = {2024},
  doi           = {10.1093/mnras/stae572},
  eprint        = {2212.03981},
  archivePrefix = {arXiv}
}

@article{Arya2023,
   title={Lognormal seminumerical simulations of the Lyman-alpha forest: comparison with full hydrodynamic simulations},
   volume={520},
   ISSN={1365-2966},
   url={http://dx.doi.org/10.1093/mnras/stad386},
   DOI={10.1093/mnras/stad386},
   number={3},
   journal={Monthly Notices of the Royal Astronomical Society},
   publisher={Oxford University Press (OUP)},
   author={Arya, Bhaskar and Choudhury, Tirthankar Roy and Paranjape, Aseem and Gaikwad, Prakash},
   year={2023},
   month=Feb, pages={4023–4036} }

@misc{Arya2024a,
      title={A modified lognormal approximation of the Lyman-alpha forest: comparison with full hydrodynamic simulations at $2\leq z\leq 2.7$}, 
      author={Bhaskar Arya and Tirthankar Roy Choudhury and Aseem Paranjape and Prakash Gaikwad},
      year={2024},
      eprint={2310.12720},
      archivePrefix={arXiv},
      primaryClass={astro-ph.CO},
      url={https://arxiv.org/abs/2310.12720}, 
}

@article{Bolton2017,
  author        = {Bolton, James S. and Puchwein, Ewald and Sijacki, Debora and Haehnelt, Martin G. and Kim, Tae-Sun and Meiksin, Avery and Regan, John A. and Viel, Matteo},
  title         = {The Sherwood simulation suite: overview and data comparisons with the Lyman-alpha forest at redshifts \(2 \leq z \leq 5\)},
  journal       = {Monthly Notices of the Royal Astronomical Society},
  volume        = {464},
  pages         = {897--914},
  year          = {2017},
  doi           = {10.1093/mnras/stw2397},
  eprint        = {1605.03462},
  archivePrefix = {arXiv}
}

@article{Lukic2015,
  author        = {Luki{\'c}, Zarija and Stark, Casey W. and Nugent, Peter and White, Martin and Meiksin, Avery A. and Almgren, Ann},
  title         = {The Lyman-alpha forest in optically thin hydrodynamical simulations},
  journal       = {Monthly Notices of the Royal Astronomical Society},
  volume        = {446},
  pages         = {3697--3724},
  year          = {2015},
  doi           = {10.1093/mnras/stu2377},
  eprint        = {1406.6361},
  archivePrefix = {arXiv}
}

@article{Schaye2015,
  author        = {Schaye, Joop and Crain, Robert A. and Bower, Richard G. and Furlong, Michelle and Schaller, Matthieu and Theuns, Tom and Dalla Vecchia, Claudio and Frenk, Carlos S. and McCarthy, Ian G. and Helly, John C. and Jenkins, Adrian and Rosas-Guevara, Y. M. and White, Simon D. M. and Baes, Maarten and Booth, C. M. and Camps, Peter and Navarro, Julio F. and Qu, Yan and Rahmati, Alireza and Sawala, Till and Thomas, Peter A. and Trayford, James},
  title         = {The EAGLE project: simulating the evolution and assembly of galaxies and their environments},
  journal       = {Monthly Notices of the Royal Astronomical Society},
  volume        = {446},
  pages         = {521--554},
  year          = {2015},
  doi           = {10.1093/mnras/stu2058},
  eprint        = {1407.7040},
  archivePrefix = {arXiv}
}

@article{Schaye2023,
  author        = {Schaye, Joop and Kugel, Roi and Schaller, Matthieu and Helly, John C. and Braspenning, Joey and Elbers, Willem and McCarthy, Ian G. and van Daalen, Marcel P. and Vandenbroucke, Bert and Frenk, Carlos S. and Kwan, Juliana and Salcido, Jaime and Bah{\'e}, Yannick M. and Borrow, Josh and Chaikin, Evgenii and Hahn, Oliver and Hu{\v{s}}ko, Filip and Jenkins, Adrian and Lacey, Cedric G. and Nobels, Folkert S. J.},
  title         = {The FLAMINGO project: cosmological hydrodynamical simulations for large-scale structure and galaxy cluster surveys},
  journal       = {Monthly Notices of the Royal Astronomical Society},
  volume        = {526},
  pages         = {4978--5020},
  year          = {2023},
  doi           = {10.1093/mnras/stad2419},
  eprint        = {2306.04024},
  archivePrefix = {arXiv}
}

@article{Chabanier2024,
  title   = {The ACCEL\textsuperscript{2} project: Simulating Lyman-\(\alpha\) forest in large-volume hydrodynamical simulations},
  author  = {Chabanier, Sol{\`e}ne and Ravoux, Corentin and Latrille, Lucas and Sexton, Jean and Armengaud, {\'E}ric and Bautista, Julian and Dumerchat, Tyann and Luki{\'c}, Zarija},
  journal = {Monthly Notices of the Royal Astronomical Society},
  year    = {2024},
  month   = oct,
  doi     = {10.1093/mnras/stae2255},
  url     = {https://doi.org/10.1093/mnras/stae2255}
}

@article{Nelson2019,
  author        = {Nelson, Dylan and Springel, Volker and Pillepich, Annalisa and Rodriguez-Gomez, Vicente and Torrey, Paul and Genel, Shy and Vogelsberger, Mark and Pakmor, R{\"u}diger and Marinacci, Federico and Weinberger, Rainer and Kelley, Luke and Lovell, Mark and Diemer, Benedikt and Hernquist, Lars},
  title         = {The IllustrisTNG simulations: public data release},
  journal       = {Computational Astrophysics and Cosmology},
  volume        = {6},
  pages         = {2},
  year          = {2019},
  doi           = {10.1186/s40668-019-0028-x},
  eprint        = {1812.05609},
  archivePrefix = {arXiv}
}

@article{Pillepich2018,
  author        = {Pillepich, Annalisa and Springel, Volker and Nelson, Dylan and Genel, Shy and Naiman, Jill and Pakmor, R{\"u}diger and Hernquist, Lars and Torrey, Paul and Vogelsberger, Mark and Weinberger, Rainer and Marinacci, Federico},
  title         = {Simulating galaxy formation with the IllustrisTNG model},
  journal       = {Monthly Notices of the Royal Astronomical Society},
  volume        = {473},
  pages         = {4077--4106},
  year          = {2018},
  doi           = {10.1093/mnras/stx2656},
  eprint        = {1703.02970},
  archivePrefix = {arXiv}
}

@article{BiDavidsen1997,
  author        = {Bi, Hongguang and Davidsen, Arthur F.},
  title         = {Evolution of Structure in the Intergalactic Medium and the Nature of the Ly-alpha Forest},
  journal       = {The Astrophysical Journal},
  volume        = {479},
  pages         = {523--542},
  year          = {1997},
  doi           = {10.1086/303908},
  eprint        = {astro-ph/9611062},
  archivePrefix = {arXiv}
}

@article{Choudhury2001a,
  author        = {Choudhury, T. Roy and Padmanabhan, T. and Srianand, R.},
  title         = {Semianalytic approach to understanding the distribution of neutral hydrogen in the universe},
  journal       = {Monthly Notices of the Royal Astronomical Society},
  volume        = {322},
  pages         = {561--575},
  year          = {2001},
  doi           = {10.1046/j.1365-8711.2001.04157.x},
  eprint        = {astro-ph/0005252},
  archivePrefix = {arXiv}
}

@article{Choudhury2001b,
  author        = {Choudhury, T. Roy and Srianand, R. and Padmanabhan, T.},
  title         = {Semianalytic approach to understanding the distribution of neutral hydrogen in the universe: comparison with observations},
  journal       = {The Astrophysical Journal},
  volume        = {559},
  pages         = {29--38},
  year          = {2001},
  doi           = {10.1086/322411},
  eprint        = {astro-ph/0012498},
  archivePrefix = {arXiv}
}

@article{FontRibera2012,
  author        = {Font-Ribera, Andreu and McDonald, Patrick and Miralda-Escud{\'e}, Jordi},
  title         = {Generating mock data sets for large-scale Lyman-alpha forest correlation measurements},
  journal       = {Journal of Cosmology and Astroparticle Physics},
  volume        = {2012},
  number        = {01},
  pages         = {001},
  year          = {2012},
  doi           = {10.1088/1475-7516/2012/01/001},
  eprint        = {1108.5606},
  archivePrefix = {arXiv}
}

@ARTICLE{Kooistra2022,
       author = {{Kooistra}, Robin and {Lee}, Khee-Gan and {Horowitz}, Benjamin},
        title = "{Constraining the Fluctuating Gunn-Peterson Approximation using Ly{\ensuremath{\alpha}} Forest Tomography at z = 2}",
      journal = {The Astrophysical Journal},
         year = 2022,
        month = oct,
       volume = {938},
       number = {2},
          eid = {123},
        pages = {123},
          doi = {10.3847/1538-4357/ac92e8},
archivePrefix = {arXiv},
       eprint = {2201.10169},
 primaryClass = {astro-ph.CO},
       adsurl = {https://ui.adsabs.harvard.edu/abs/2022ApJ...938..123K}
}

@ARTICLE{Sinigaglia2024,
       author = {{Sinigaglia}, F. and {Kitaura}, F.-S. and {Nagamine}, K. and {Oku}, Y. and {Balaguera-Antol{\'\i}nez}, A.},
        title = "{Field-level Lyman-{\ensuremath{\alpha}} forest modeling in redshift space via augmented nonlocal Fluctuating Gunn-Peterson Approximation}",
      journal = {Astronomy and Astrophysics},
         year = 2024,
        month = feb,
       volume = {682},
          eid = {A21},
        pages = {A21},
          doi = {10.1051/0004-6361/202346931},
archivePrefix = {arXiv},
       eprint = {2305.10428},
 primaryClass = {astro-ph.CO},
       adsurl = {https://ui.adsabs.harvard.edu/abs/2024A&A...682A..21S}
}

@article{Etourneau2024,
  author        = {Etourneau, Thomas and Le Goff, Jean-Marc and Rich, James and Tan, Ting and Cuceu, Andrei and others},
  title         = {Mock data sets for the eBOSS and DESI Lyman-alpha forest surveys},
  journal       = {Journal of Cosmology and Astroparticle Physics},
  volume        = {2024},
  number        = {02},
  pages         = {003},
  year          = {2024},
  doi           = {10.1088/1475-7516/2024/02/003},
  eprint        = {2310.18996},
  archivePrefix = {arXiv}
}

@article{Peirani2014,
  author        = {Peirani, S{\'e}bastien and Weinberg, David H. and Colombi, St{\'e}phane and Blaizot, J{\'e}r{\'e}my and Dubois, Yohan and Pichon, Christophe},
  title         = {LyMAS: Predicting Large-Scale Lyman-alpha Forest Statistics from the Dark Matter Density Field},
  journal       = {The Astrophysical Journal},
  volume        = {784},
  pages         = {11},
  year          = {2014},
  doi           = {10.1088/0004-637X/784/1/11},
  eprint        = {1306.1533},
  archivePrefix = {arXiv}
}

@article{Peirani2022,
  author        = {Peirani, S. and Prunet, S. and Colombi, S. and Pichon, C. and Weinberg, D. H. and Laigle, C. and Lavaux, G. and Dubois, Y. and Devriendt, J.},
  title         = {LyMAS reloaded: improving the predictions of the large-scale Lyman-alpha forest statistics from dark matter density and velocity fields},
  journal       = {Monthly Notices of the Royal Astronomical Society},
  volume        = {514},
  pages         = {3222--3236},
  year          = {2022},
  doi           = {10.1093/mnras/stac1465},
  eprint        = {2204.06365},
  archivePrefix = {arXiv}
}

@article{Sinigaglia2022,
  author        = {Sinigaglia, Francesco and Kitaura, Francisco-Shu and Balaguera-Antol{\'i}nez, Andr{\'e}s and Shimizu, Ikkoh and Nagamine, Kentaro and S{\'a}nchez-Benavente, Manuel and Ata, Metin},
  title         = {Mapping the Three-dimensional Ly-alpha Forest Large-scale Structure in Real and Redshift Space},
  journal       = {The Astrophysical Journal},
  volume        = {927},
  pages         = {230},
  year          = {2022},
  doi           = {10.3847/1538-4357/ac4bde},
  eprint        = {2107.07917},
  archivePrefix = {arXiv}
}

@misc{Arya2024b,
      title={Covariance matrices for the Lyman-alpha forest using the lognormal approximation}, 
      author={Bhaskar Arya and Aseem Paranjape and Tirthankar Roy Choudhury},
      year={2024},
      eprint={2310.16464},
      archivePrefix={arXiv},
      primaryClass={astro-ph.CO},
      url={https://arxiv.org/abs/2310.16464}, 
}

@article{Bautista2015,
  author        = {Bautista, Julian E. and Bailey, Stephen and Font-Ribera, Andreu and Slosar, An{\v z}e and Lee, Khee-Gan and Pieri, Matthew M. and Busca, Nicol{\'a}s G. and Miralda-Escud{\'e}, Jordi and Palanque-Delabrouille, Nathalie and Rich, James and others},
  title         = {Mock Quasar-Lyman-alpha Forest Data-sets for the SDSS-III Baryon Oscillation Spectroscopic Survey},
  journal       = {Journal of Cosmology and Astroparticle Physics},
  volume        = {2015},
  number        = {05},
  pages         = {060},
  year          = {2015},
  doi           = {10.1088/1475-7516/2015/05/060},
  eprint        = {1412.0658},
  archivePrefix = {arXiv}
}

@article{Farr2020,
  author        = {Farr, James and Font-Ribera, Andreu and du Mas des Bourboux, H{\'e}lion and Mu{\~n}oz-Guti{\'e}rrez, Andrea and Sanchez, Francisco Javier and Pontzen, Andrew and Gonz{\'a}lez-Morales, Alma Xochitl and Alonso, David and Brooks, David and Doel, Peter and Etourneau, Thomas and Guy, Julien and Le Goff, Jean-Marc and de la Macorra, Axel and Palanque-Delabrouille, Nathalie and P{\'e}rez-R{\`a}fols, Ignasi and Rich, James and Slosar, An{\v z}e and Tarle, Gregory and Yutong, Duan and Zhang, Kai},
  title         = {LyaCoLoRe: Synthetic Datasets for Current and Future Lyman-alpha Forest BAO Surveys},
  journal       = {Journal of Cosmology and Astroparticle Physics},
  volume        = {2020},
  number        = {03},
  pages         = {068},
  year          = {2020},
  doi           = {10.1088/1475-7516/2020/03/068},
  eprint        = {1912.02763},
  archivePrefix = {arXiv}
}

@inproceedings{Ronneberger2015,
  author        = {Ronneberger, Olaf and Fischer, Philipp and Brox, Thomas},
  title         = {U-Net: Convolutional Networks for Biomedical Image Segmentation},
  booktitle     = {Medical Image Computing and Computer-Assisted Intervention -- MICCAI 2015},
  pages         = {234--241},
  year          = {2015},
  publisher     = {Springer},
  doi           = {10.1007/978-3-319-24574-4_28},
  eprint        = {1505.04597},
  archivePrefix = {arXiv}
}

@inproceedings{Cicek2016,
  author        = {{\c{C}}i{\c{c}}ek, {\"O}zg{\"u}n and Abdulkadir, Ahmed and Lienkamp, Soeren S. and Brox, Thomas and Ronneberger, Olaf},
  title         = {3D U-Net: Learning Dense Volumetric Segmentation from Sparse Annotation},
  booktitle     = {Medical Image Computing and Computer-Assisted Intervention -- MICCAI 2016},
  pages         = {424--432},
  year          = {2016},
  publisher     = {Springer},
  doi           = {10.1007/978-3-319-46723-8_49},
  eprint        = {1606.06650},
  archivePrefix = {arXiv}
}

@article{Maitra2025,
  author        = {Maitra, Soumak and Viel, Matteo and Kulkarni, Girish},
  title         = {DeepCHART: Mapping the 3D dark matter density field from Ly\(\alpha\) forest surveys using deep learning},
  journal       = {arXiv e-prints},
  pages         = {arXiv:2507.00135},
  year          = {2025},
  eprint        = {2507.00135},
  archivePrefix = {arXiv}
}

@article{Harrington:2021srm,
    author = "Harrington, Peter and Mustafa, Mustafa and Dornfest, Max and Horowitz, Benjamin and Luki{\'c}, Zarija",
    title = "{Fast, High-fidelity Ly{\ensuremath{\alpha}} Forests with Convolutional Neural Networks}",
    eprint = "2106.12662",
    archivePrefix = "arXiv",
    primaryClass = "astro-ph.CO",
    doi = "10.3847/1538-4357/ac5faa",
    journal = "Astrophys. J.",
    volume = "929",
    number = "2",
    pages = "160",
    year = "2022"
}

@article{Horowitz2021olb,
    author = "Horowitz, Benjamin and Dornfest, Max and Luki{\'c}, Zarija and Harrington, Peter",
    title = "{hyphy: Deep Generative Conditional Posterior Mapping of Hydrodynamical Physics}",
    eprint = "2106.12675",
    archivePrefix = "arXiv",
    primaryClass = "astro-ph.CO",
    doi = "10.3847/1538-4357/ac9ea7",
    journal = "Astrophys. J.",
    volume = "941",
    number = "1",
    pages = "42",
    year = "2022"
}

@misc{Boonkongkird2023,
      title={LyAl-Net: A high-efficiency Lyman-{\ensuremath{\alpha}} forest simulation with a neural network}, 
      author={Chotipan Boonkongkird and Guilhem Lavaux and Sebastien Peirani and Yohan Dubois and Natalia Porqueres and Eleni Tsaprazi},
      year={2023},
      eprint={2303.17939},
      archivePrefix={arXiv},
      primaryClass={astro-ph.CO},
      url={https://arxiv.org/abs/2303.17939}, 
}

@misc{jacobus2023,
      title={Reconstructing Lyman-{\ensuremath{\alpha}} Fields from Low-Resolution Hydrodynamical Simulations with Deep Learning}, 
      author={Cooper Jacobus and Peter Harrington and Zarija Lukić},
      year={2023},
      eprint={2308.02637},
      archivePrefix={arXiv},
      primaryClass={astro-ph.CO},
      url={https://arxiv.org/abs/2308.02637}, 
}

@ARTICLE{Jacobus2025,
       author = {{Jacobus}, Cooper and {de Belsunce}, Roger and {Chabanier}, Sol{\`e}ne and {Harrington}, Peter and {Emberson}, J.~D. and {Luki{\'c}}, Zarija and {Habib}, Salman},
        title = "{A Gigaparsec-scale Hydrodynamic Volume Reconstructed with Deep Learning}",
      journal = {apj},
         year = 2025,
        month = nov,
       volume = {993},
       number = {2},
          eid = {243},
        pages = {243},
          doi = {10.3847/1538-4357/ae0720},
archivePrefix = {arXiv},
       eprint = {2411.16920},
 primaryClass = {astro-ph.CO},
       adsurl = {https://ui.adsabs.harvard.edu/abs/2025ApJ...993..243J}
}

@ARTICLE{Springel2010,
       author = {{Springel}, Volker},
        title = "{E pur si muove: Galilean-invariant cosmological hydrodynamical simulations on a moving mesh}",
      journal = {mnras},
         year = 2010,
        month = jan,
       volume = {401},
       number = {2},
        pages = {791-851},
          doi = {10.1111/j.1365-2966.2009.15715.x},
archivePrefix = {arXiv},
       eprint = {0901.4107},
 primaryClass = {astro-ph.CO},
       adsurl = {https://ui.adsabs.harvard.edu/abs/2010MNRAS.401..791S}
}

@ARTICLE{Schaller2024,
       author = {{Schaller}, Matthieu and {Borrow}, Josh and {Draper}, Peter W. and {Ivkovic}, Mladen and {McAlpine}, Stuart and {Vandenbroucke}, Bert and {Bah{\'e}}, Yannick and {Chaikin}, Evgenii and {Chalk}, Aidan B.~G. and {Chan}, Tsang Keung and {Correa}, Camila and {van Daalen}, Marcel and {Elbers}, Willem and {Gonnet}, Pedro and {Hausammann}, Lo{\"\i}c and {Helly}, John and {Hu{\v{s}}ko}, Filip and {Kegerreis}, Jacob A. and {Nobels}, Folkert S.~J. and {Ploeckinger}, Sylvia and {Revaz}, Yves and {Roper}, William J. and {Ruiz-Bonilla}, Sergio and {Sandnes}, Thomas D. and {Uyttenhove}, Yolan and {Willis}, James S. and {Xiang}, Zhen},
        title = "{SWIFT: A modern highly-parallel gravity and smoothed particle hydrodynamics solver for astrophysical and cosmological applications}",
      journal = {mnras},
         year = 2024,
        month = may,
       volume = {530},
       number = {2},
        pages = {2378-2419},
          doi = {10.1093/mnras/stae922},
archivePrefix = {arXiv},
       eprint = {2305.13380},
 primaryClass = {astro-ph.IM},
       adsurl = {https://ui.adsabs.harvard.edu/abs/2024MNRAS.530.2378S}
}

@article{Meiksin2009,
    author  = {Meiksin, Avery A.},
    title   = {The Physics of the Intergalactic Medium},
    journal = {Reviews of Modern Physics},
    volume  = {81},
    number  = {4},
    pages   = {1405--1469},
    year    = {2009},
    doi     = {10.1103/RevModPhys.81.1405},
    eprint  = {0711.3358},
    archivePrefix = {arXiv},
    primaryClass  = {astro-ph}
}

@article{McQuinn2016,
    author  = {McQuinn, Matthew},
    title   = {The Evolution of the Intergalactic Medium},
    journal = {Annual Review of Astronomy and Astrophysics},
    volume  = {54},
    pages   = {313--362},
    year    = {2016},
    doi     = {10.1146/annurev-astro-082214-122355},
    eprint  = {1512.00086},
    archivePrefix = {arXiv},
    primaryClass  = {astro-ph.CO}
}

@article{Seljak2006,
    author  = {Seljak, Uro{\v s} and Slosar, An{\v z}e and McDonald, Patrick},
    title   = {Cosmological Parameters from Combining the
               {Lyman-alpha} Forest with {CMB}, Galaxy Clustering,
               and {SN} Constraints},
    journal = {Journal of Cosmology and Astroparticle Physics},
    volume  = {2006},
    number  = {10},
    pages   = {014},
    year    = {2006},
    doi     = {10.1088/1475-7516/2006/10/014},
    eprint  = {astro-ph/0604335},
    archivePrefix = {arXiv}
}

@article{Viel2010,
    author  = {Viel, Matteo and Haehnelt, Martin G. and Springel, Volker},
    title   = {The Effect of Neutrinos on the Matter Distribution
               as Probed by the Intergalactic Medium},
    journal = {Journal of Cosmology and Astroparticle Physics},
    volume  = {2010},
    number  = {06},
    pages   = {015},
    year    = {2010},
    doi     = {10.1088/1475-7516/2010/06/015},
    eprint  = {1003.2422},
    archivePrefix = {arXiv},
    primaryClass  = {astro-ph.CO}
}

@article{Yeche2017,
    author  = {Y{\`e}che, Christophe and Palanque-Delabrouille, Nathalie
               and Baur, Julien and du Mas des Bourboux, H{\'e}lion},
    title   = {Constraints on Neutrino Masses from
               {Lyman-alpha} Forest Power Spectrum with
               {BOSS} and {XQ-100}},
    journal = {Journal of Cosmology and Astroparticle Physics},
    volume  = {2017},
    number  = {06},
    pages   = {047},
    year    = {2017},
    doi     = {10.1088/1475-7516/2017/06/047},
    eprint  = {1702.03314},
    archivePrefix = {arXiv},
    primaryClass  = {astro-ph.CO}
}

@article{Irsic2017WDM,
    author  = {Ir{\v s}i{\v c}, Vid and others},
    title   = {New Constraints on the Free-Streaming of Warm Dark Matter
               from Intermediate and Small Scale
               {Lyman-alpha} Forest Data},
    journal = {Physical Review D},
    volume  = {96},
    number  = {2},
    pages   = {023522},
    year    = {2017},
    doi     = {10.1103/PhysRevD.96.023522}
}

@article{Irsic2017FDM,
    author  = {Ir{\v s}i{\v c}, Vid and Viel, Matteo
               and Haehnelt, Martin G. and Bolton, James S.
               and Becker, George D.},
    title   = {First Constraints on Fuzzy Dark Matter from
               {Lyman-alpha} Forest Data and Hydrodynamical Simulations},
    journal = {Physical Review Letters},
    volume  = {119},
    number  = {3},
    pages   = {031302},
    year    = {2017},
    doi     = {10.1103/PhysRevLett.119.031302},
    eprint  = {1703.04683},
    archivePrefix = {arXiv},
    primaryClass  = {astro-ph.CO}
}

@article{Murgia2018,
    author  = {Murgia, Riccardo and Ir{\v s}i{\v c}, Vid
               and Viel, Matteo},
    title   = {Novel Constraints on Noncold, Nonthermal Dark Matter
               from {Lyman-alpha} Forest Data},
    journal = {Physical Review D},
    volume  = {98},
    number  = {8},
    pages   = {083540},
    year    = {2018},
    doi     = {10.1103/PhysRevD.98.083540},
    eprint  = {1806.08371},
    archivePrefix = {arXiv},
    primaryClass  = {astro-ph.CO}
}

@article{RogersPeiris2021,
    author  = {Rogers, Keir K. and Peiris, Hiranya V.},
    title   = {Strong Bound on Canonical Ultralight Axion Dark Matter
               from the {Lyman-alpha} Forest},
    journal = {Physical Review Letters},
    volume  = {126},
    number  = {7},
    pages   = {071302},
    year    = {2021},
    doi     = {10.1103/PhysRevLett.126.071302},
    eprint  = {2007.12705},
    archivePrefix = {arXiv},
    primaryClass  = {astro-ph.CO}
}

@article{Baur2017Sterile,
    author  = {Baur, Julien and Palanque-Delabrouille, Nathalie
               and Y{\`e}che, Christophe and Boyarsky, Alexey
               and Ruchayskiy, Oleg and Armengaud, Eric
               and Lesgourgues, Julien},
    title   = {Constraints from {Ly-alpha} Forests on Non-Thermal
               Dark Matter Including Resonantly Produced
               Sterile Neutrinos},
    journal = {Journal of Cosmology and Astroparticle Physics},
    volume  = {2017},
    number  = {12},
    pages   = {013},
    year    = {2017},
    doi     = {10.1088/1475-7516/2017/12/013},
    eprint  = {1706.03118},
    archivePrefix = {arXiv},
    primaryClass  = {astro-ph.CO}
}

\end{document}